\documentclass{emulateapj}

\usepackage{natbib}
\newcommand{\msol}{{\rm M}_{\odot}}
\newcommand{\rsol}{{\rm R}_{\odot}}
\newcommand{\lsol}{{\rm L}_{\odot}}
\newcommand{\kms}{\rm km\ s^{-1}}
\newcommand{\oneday}{\rm day}
\newcommand{\days}{\rm days}

\newcommand{\theobject}{LSPM~J1112+7626}
\newcommand{\theimag}{12.14 \pm 0.05}
\newcommand{\theperiod}{41.03236 \pm 0.00002}
\newcommand{\thema}{0.395 \pm 0.002}
\newcommand{\themb}{0.275 \pm 0.001}
\newcommand{\theecc}{0.239 \pm 0.002}
\newcommand{\theooeperiod}{65}
\newcommand{\theooeperiodsf}{64.8}
\newcommand{\theinflation}{3.8^{+0.9}_{-0.5}\%}

\newcommand{\ibess}{I_{\rm Bessell}}
\newcommand{\icous}{I_C}
\newcommand{\imear}{I_{\rm MEarth}}

\newcommand{\teff}{T_{\rm eff}}

\begin{document}
\shorttitle{\theobject}
\shortauthors{Irwin et al.}

\title{\theobject: detection of a 41-day M-dwarf eclipsing binary
  from the MEarth transit survey}

\author{Jonathan~M.~Irwin, Samuel~N.~Quinn, Zachory~K.~Berta,
  David~W.~Latham, Guillermo~Torres, Christopher~J.~Burke,
  David~Charbonneau, Jason~Dittmann, Gilbert~A.~Esquerdo, and
  Robert~P.~Stefanik} 
\affil{Harvard-Smithsonian Center for Astrophysics, 60 Garden St.,
  Cambridge, MA 02138, USA}
\email{jirwin -at- cfa -dot- harvard -dot- edu}

\author{Arto~Oksanen}
\affil{Hankasalmi Observatory, Jyv\"askyl\"an Sirius ry, Vertaalantie 419,
FI-40270 Palokka, Finland}

\author{Lars~A.~Buchhave}
\affil{Niels Bohr Institute, University of Copenhagen, DK-2100 Copenhagen, Denmark}
\affil{Centre for Star and Planet Formation, Natural History Museum of Denmark,
University of Copenhagen, DK-1350 Copenhagen, Denmark}

\author{Philip~Nutzman}
\affil{Department of Astronomy and Astrophysics, University of California, Santa Cruz, CA 95064, USA}

\author{Perry~Berlind, Michael~L.~Calkins and Emilio~E.~Falco}
\affil{Fred Lawrence Whipple Observatory, Smithsonian Astrophysical Observatory, 670 Mount Hopkins Road,
  Amado, AZ 85645, USA}

\begin{abstract}
We report the detection of eclipses in \theobject, which we find to be
a moderately bright ($\icous = \theimag$) very low-mass binary system
with an orbital period of $\theperiod\ \days$, and component masses
$M_1 = \thema\ \msol$ and $M_2 = \themb\ \msol$ in an eccentric ($e =
\theecc$) orbit.  A $\theooeperiod\ \oneday$ out of eclipse modulation
of approximately $2\%$ peak-to-peak amplitude is seen in $I$-band,
which is probably due to rotational modulation of photospheric spots
on one of the binary components.  This paper presents the discovery
and characterization of the object, including radial velocities
sufficient to determine both component masses to better than $1\%$
precision, and a photometric solution.  We find that the sum of the
component radii, which is much better-determined than the individual
radii, is inflated by $\theinflation$ compared to the theoretical
model predictions, depending on the age and metallicity assumed.
These results demonstrate that the difficulties in reproducing
observed M-dwarf eclipsing binary radii with theoretical models are
not confined to systems with very short orbital periods.  This object
promises to be a fruitful testing ground for the hypothesized link
between inflated radii in M-dwarfs and activity.
\end{abstract}


\section{Introduction}
\label{intro_sect}

Detached, double-lined eclipsing binaries (EBs) provide a largely
model-independent means to precisely and accurately measure
fundamental stellar properties, particularly masses and radii.  In the
best-observed systems the precision of these can be at the $< 1$ per
cent level, and thus place stringent constraints on stellar evolution
models (e.g. \citealt{1991A&ARv...3...91A,2010A&ARv..18...67T}).

Despite the ubiquity of M-dwarfs in the solar neighborhood, their
fundamental properties are poorly understood.  This is particularly
the case below $0.35\ \msol$, the boundary at which field stars are
thought to become fully convective
(e.g. \citealt{1997A&A...327.1039C}).  Furthermore, observations of
many of the best-characterized M-dwarf EBs indicate significant
discrepancies with the stellar models.

The components of CM~Dra, the best characterized double-lined system
in the fully-convective mass range
\citep{1967ApJ...148..911E,1977ApJ...218..444L,1996ApJ...456..356M,2009ApJ...691.1400M},
have radii approximately $5-7\%$ larger than predicted by theoretical
stellar evolution models.  The observation of inflated radii for
M-dwarf binaries at the $5-10\%$ level, and effective temperatures
$3-5\%$ lower than the models predict, is nearly ubiquitous among the
best-characterized objects
(e.g. \citealt{2010A&ARv..18...67T}, and references therein;
\citealt{2010ApJ...718..502M,2011ApJ...728...48K}), although it
is interesting to note that \citet{2011Sci...331..562C} find a smaller
inflation for KOI-126B and C, members of a fascinating system
containing a short-period M-dwarf binary orbiting a K-star, where the
system undergoes mutual eclipses allowing a precise determination of
the radii of both M-dwarfs.

Two main hypotheses have been advanced to explain the inflated radii
in low-mass eclipsing binaries: metallicity
\citep{2006ApJ...644..475B}, and the effect of magnetic activity (the
``activity hypothesis'', e.g.
\citealt{2005ApJ...631.1120L,2006Ap&SS.304...89R,2007A&A...472L..17C}).

As discussed by \citet{2007ApJ...660..732L}, one of the main
difficulties with metallicity as an explanation is that current models
of M-dwarfs yield radii differing only by a very small amount (an
increase of approximately $3\%$) between $[{\rm M/H}] = -0.5$ to $0$,
the range spanned by the available \citet{1998A&A...337..403B} models.
Unless the EB sample contains many objects of very high metallicity,
this effect does not seem large enough.\footnote{As discussed by
\citet{2007ApJ...660..732L}, it is possible the effect on the radius
is suppressed by a missing source of opacity in the models.  Such
effects are known to exist as the same models fail to reproduce the
observed optical spectra and colors of M-dwarfs
\citep{1998A&A...337..403B}.}

Observationally, the difficulty of determining metallicities for
M-dwarfs (e.g.
\citealt{2005A&A...442..635B,2006PASP..118..218W,2006ApJ...652.1604B})
renders the metallicity hypothesis difficult to test in practice, and
this is further exacerbated by the double lines and rapid rotation in
most M-dwarf eclipsing binary spectra.\footnote{We thank the referee
for bringing this point to our attention.}  Nonetheless, it is
interesting to note that CM~Dra is thought to be metal poor
\citep{1997MNRAS.291..780V,2002MNRAS.329..290V}, which would increase
the size of the radius discrepancy for this object, rather than
decreasing it, and more generally, this is expected to also be true of
other eclipsing binaries in the sample, assuming they follow the
metallicity distribution of field stars
(e.g. \citealt{1989AJ.....97..423C}).

Given these lines of reasoning, it seems unlikely that metallicity is
the only explanation for the inflated radii (although it probably
plays a role, particularly in creating scatter in the effective
temperatures).  This has led to the ``activity hypothesis'' becoming
the preferred explanation in the literature for the inflation.  It is
motivated by noting that nearly all EB systems have short orbital
periods.  For example, below $0.35\ \msol$, the longest period
system is 1RXS~J154727.5+450803 at $3.55\ {\rm days}$
\citep{2011AJ....141..166H}.  At such short periods, the components
are expected to have been tidally synchronized and the orbits
circularized by field ages.  The effect of the companion and tidal
locking is likely to significantly increase the activity levels.  The
available observational evidence supports this, with the eclipsing
binaries below $0.35\ \msol$ all showing signs of high activity and
tidal effects including synchronous, rapid rotation, large amplitude
out of eclipse modulations, H$\alpha$ emission, X-ray emission, and in
most cases, circular orbits.

Magnetic activity could have an effect either by inhibiting convection
\citep{2001ApJ...559..353M,2007A&A...472L..17C}, or due to reduced
heat flux resulting from cool photospheric spots.  Of these,
\citet{2007A&A...472L..17C} find the structure of objects below $0.35\
\msol$ is almost independent of variations in the mixing length
parameter, which they use to implement inhibition of convection, so
the latter possibility (inhibition of radiation due to the effect of
spots) seems more likely the explanation in the mass domain of
particular interest here.  It is important to note that spots also
influence the solution of eclipsing binary light curves. The
possibility of such systematic errors in the observations is discussed
by \citet{2010ApJ...718..502M}, and we shall return to it later in the
analysis.

An obvious way to test the activity hypothesis is to measure radii for
components of eclipsing binaries with long orbital periods, where
tides are unimportant, and the stars should rotate close to the rates
expected for single stars in the absence of a binary companion.
Unfortunately, such an endeavor is difficult due to the reduced
geometric alignment probability and limited availability of
long-duration photometric monitoring of M-dwarfs at rapid cadences,
where the majority of wide-field variability surveys are carried out
in visible bandpasses, often yielding poor signal to noise ratios for
M-dwarfs, which are faint at these wavelengths.

The longest-period M-dwarf EB with a published orbit is T-Lyr1-17236
\citep{2008ApJ...687.1253D}, which has an $8.4\ \oneday$ orbital
period and components of $0.68$ and $0.52\ \msol$.  The radii are
indeed consistent with the theoretical models, but the observational
uncertainties are still quite large in the discovery paper.  This
object is currently being observed by the NASA Kepler mission, so a
more precise solution should be forthcoming.

In light of the difficulty of identifying longer-period EBs, several
authors have attempted to test the activity hypothesis using the
existing EB sample to search for correlations between e.g. orbital
period or activity measures and radius inflation
\citep{2007ApJ...660..732L,2011ApJ...728...48K}.  While there does
appear to be evidence for such a correlation above $0.35\ \msol$,
below this mass the present sample is limited by small number
statistics and the narrow range of orbital periods spanned.  The
Kepler mission appears to have identified a large number of very
long-period EB systems (\citealt{2011AJ....141...83P}; see also
\citealt{2011AJ....141...78C} where more detailed solutions using the
Kepler photometry are presented for low-mass objects from this
sample), but we are not aware of published dynamical masses and
model-independent radii for these at the present time, and many of the
M-dwarfs (especially below $0.35\ \msol$) are secondaries in systems
with mass ratios much less than unity, where detecting the secondary
spectrum to measure radial velocities may be challenging.

An alternative method to determine M-dwarf radii is by making
interferometric measurements of angular diameters for (usually
isolated) stars.  Combined with trigonometric parallaxes and
minor assumptions about the stellar limb darkening, the physical
radius of the star can be inferred in a model-independent fashion.
The difficulty with this method comes in estimating the stellar mass.
Dynamical measurements are usually not available, so one has to resort
to inferring mass from other measurable stellar properties,
necessarily with larger uncertainties.  Typically, absolute magnitude
in near-infrared passbands is used, which is found to correlate well
with stellar mass and has a relatively small scatter.  The empirical
polynomials of \citet{2000A&A...364..217D} are the most widely used,
although there is some argument in the literature as to how accurately
these predict the mass, especially over a range of metallicity and
spectral type (the uncertainties are probably $\ge 10\%$).  These
concerns combined with small sample sizes mean interferometry 
results have thus far remained somewhat ambiguous
(e.g. \citealt{2009A&A...505..205D,2010arXiv1012.0542B}, and
references therein).

In this contribution, we present the discovery of eclipses in
LSPM~J1112+7626, a nearby double-lined M-dwarf system with an
extremely long orbital period (compared to existing well-characterized
M-dwarf EBs) of approximately $41\ \days$.  At this large orbital
separation the system is not expected to have experienced significant
tidal effects (e.g. \citealt{1977A&A....57..383Z}), and is predicted
to be neither synchronized nor circularized, as verified by the
non-circular orbit and apparent non-synchronized spin and orbit for
one of the components as we shall describe.  In the absence of strong
tidal effects, it is expected that the angular momentum evolution of
each component has not been perturbed significantly by the binary
companion, and therefore the stars should possess magnetic properties
more representative of single M-dwarfs.  In this contribution, we
proceed to derive a high-quality spectroscopic orbit and a 
photometric solution for the system.

\section{Observations and data reduction}

\subsection{Initial detection}
\label{obs_detect_sect}

\theobject\ was targeted as part of routine operation of the MEarth
transit survey \citep{2008PASP..120..317N,2009IAUS..253...37I} during
the 2009-2010 season, with the first observations taken on UT 2009
December 2.  Exposure times for the \theobject\ field were $27\ {\rm
  s}$, taking three exposures per visit (the total exposure times are
tailored for each target to achieve sensitivity to a particular planet
size for the assumed stellar radius), with visits at a cadence of $20\
{\rm minutes}$.

Eclipses were first detected in \theobject\ on UT 2010 April 3 (this
was a primary eclipse), and subsequently confirmed on UT 2010 April
28th by observing the majority of a secondary eclipse egress.  Both
eclipses were detected by an experimental automated real-time
eclipse detection and followup system we began to trial during the
2009-2010 season (this will be described in detail in a future
publication when it has matured).  Confident that the object was an
eclipsing binary, but still with an ambiguous orbital period, we
immediately commenced radial velocity monitoring.  The radial
velocities ultimately determined which of the possible orbital periods
was correct, in conjunction with a further eclipse (which was also a
secondary) observed on UT 2010 June 8 with MEarth and the Clay
telescope (see \S \ref{obs_sec_sect}).

\subsection{Secondary eclipses}
\label{obs_sec_sect}

By the time the orbital ephemeris was well-known, the observing season
on \theobject\ from Arizona was ending, and the weather had begun to
deteriorate due to the arrival of the monsoon.  We therefore took
advantage of the high declination of our target and commenced an
observing program to target the secondary eclipses using the Harvard
University $0.4$m Landon T. Clay telescope, which is located on
the roof of the Science Center on the Harvard campus near Harvard
Square, Cambridge, Massachusetts.  This instrument is normally used
for undergraduate teaching.  It is equipped with an e2v CCD47-10 $1024
\times 1024$ pixel back-illuminated CCD packaged by Apogee Instruments
(model AP47), which was operated binned $2 \times 2$ for a final pixel
scale of $\approx 1\farcs59$ per pixel to reduce overheads.

The location of the telescope is extremely light polluted, and
combined with the red colors of our target, we decided to use the
$I$-band filter to mitigate the effects of sky background and
atmospheric extinction.  The filter available was built to the
prescription of \citet{1990PASP..102.1181B} using Schott RG9 glass to
define the bandpass, and shall hereafter be referred to as $\ibess$.
While this filter was intended to approximate the Cousins $I$ passband
(hereafter $\icous$), the combination of the glass filter and CCD
quantum efficiency yields a system response with significant
sensitivity at very red wavelengths not present in the original
Cousins passband, and thus is a poor approximation to the Cousins
filter for very red stars such as M-dwarfs.  We shall return to this
issue in \S \ref{mod_band_sect}.

Data were successfully gathered using the Clay telescope for the
secondary eclipses on UT 2010 June 8, October 9, November 19, and
December 30th.  The entire eclipse was visible for all but the June
8th event, although varying fractions of these were lost due to
clouds.  Exposure times were adjusted to keep the peak counts in the
target below $30\,000$ to avoid saturation, and ranged from $30-60\
{\rm s}$.  The typical cadence was $50\ {\rm s}$ and the FWHM of the
stellar images was $2-3$ binned pixels.  No guiding was available, so
the target was re-centered manually approximately every $20$ exposures
to attempt to keep the drift below $5$ pixels.

Data were reduced and light curves produced using standard procedures
for CCD data, following the method outlined in
\citet{2007MNRAS.375.1449I}.  2-D bias frames were subtracted as there
was some bias structure evident, and we used the overscan region to
correct for any bias level variations.  Dark frames of exposure
matched to the target frames were subtracted to correct for the
(significant) dark current and hot pixels at the $-20$C operating
temperature of the CCD, and the data were flat-fielded using twilight
flat-fields in the usual way.  No defringing or absolute calibration
of the photometry were attempted. Fringing was clearly visible in the
images, but at a low amplitude ($3\%$ peak-to-peak) and should not
have a significant effect on the photometry due to the stabilized
telescope pointing.

We present the full light curve data-set used in this paper in Table
\ref{lc-data}, and times of minimum light for the eclipses where a
clear minimum was seen in Table \ref{minlight}.

\begin{deluxetable*}{lrrrrrrrrrrr}
\tabletypesize{\normalsize}
\tablecaption{\label{lc-data} Light curve data.}
\tablecolumns{12}

\tablehead{
\colhead{ID\tablenotemark{a}} & \colhead{Orbit\tablenotemark{b}} &
\colhead{HJD\_UTC} & \colhead{$I$} & \colhead{$\sigma(I)$} & \colhead{$\Delta$
  mag\tablenotemark{c}} & \colhead{FWHM\tablenotemark{d}} & \colhead{Airmass} &
\colhead{MeridFlag\tablenotemark{e}} & \colhead{$x$} & \colhead{$y$} &
\colhead{CM\tablenotemark{f}} \\
 & & \colhead{(days)} & \colhead{(mag)} & \colhead{(mag)} &
\colhead{(mag)} & \colhead{(pix)} & & & \colhead{(pix)} &
\colhead{(pix)} & \colhead{(mag)}
}

\startdata
C &$1$ &$2455355.573517$ &$-0.0224$ &$0.0073$ &$ 0.000$ &$ 2.10$ &$1.2814$ &$0$ &$ 100.427$ &$ 378.595$ &$0.0000$ \\
C &$1$ &$2455355.574050$ &$-0.0168$ &$0.0077$ &$-0.002$ &$ 2.19$ &$1.2820$ &$0$ &$ 100.715$ &$ 378.567$ &$0.0000$ \\
C &$1$ &$2455355.574547$ &$-0.0161$ &$0.0079$ &$-0.000$ &$ 2.18$ &$1.2827$ &$0$ &$ 100.936$ &$ 378.442$ &$0.0000$ \\
C &$1$ &$2455355.575091$ &$-0.0176$ &$0.0079$ &$-0.033$ &$ 2.26$ &$1.2833$ &$0$ &$ 100.904$ &$ 378.255$ &$0.0000$ \\
C &$1$ &$2455355.575624$ &$-0.0165$ &$0.0081$ &$-0.010$ &$ 2.20$ &$1.2840$ &$0$ &$ 100.900$ &$ 378.023$ &$0.0000$ \\
\enddata

\tablenotetext{a}{Data-set identifier.
          C: Clay telescope secondary eclipse (magnitudes for each eclipse /
             night were separately normalized to zero median)
          H: Hankasalmi primary eclipse (normalized to zero median as above)
          O: MEarth out of eclipse (magnitudes in the instrumental system)
          P: MEarth primary eclipse (715nm long-pass, normalized as above)
          S: MEarth secondary eclipse (normalized as above)}
\tablenotetext{b}{Orbit number (from zero at $T_0$).  Given only for the
          data-sets containing eclipses.  These are the numbers used
          in the figures showing the light curves.}
\tablenotetext{c}{Photometric zero-point correction applied by the
          differential photometry procedure.  Note: this has already
          been included in the $I$ column and is for reference
          purposes only (e.g. detecting frames with large light losses
          due to clouds).}
\tablenotetext{d}{FWHM of the stellar images on the frame.  Zero in
          cases where the FWHM could not be reliably estimated.}
\tablenotetext{e}{Flag for detecting ``meridian flip'' on German
          equatorial mountings.  Used only for the MEarth data, where
          the flag is 0 when the telescope was observing in the usual
          orientation for the east side and 1 for the west side of the
          meridian.  Note that there is not a one-to-one relation with
          the sign of the hour angle because the MEarth mounts can
          track approximately 5 degrees across the meridian before
          needing to change sides of the pier, so this flag can
          sometimes be 0 even for (small) positive hour angle.}
\tablenotetext{f}{For the MEarth data only.  Gives the ``common mode''
          interpolated to the Julian date of the exposure.  Derived
          from the average differential magnitude of all the M-dwarfs
          observed by all 8 MEarth telescopes in a given time
          interval.  This should be scaled and subtracted from the $I$
          column to correct for the suspected variations in the MEarth
          bandpass with humidity and temperature (see text).}

\tablecomments{Table \ref{lc-data} is published in its
  entirety in the electronic edition of the Astrophysical Journal.  A
  portion is shown here for guidance regarding its form and content.}

\end{deluxetable*}

\begin{deluxetable}{lrrll}
\tabletypesize{\normalsize}
\tablecaption{\label{minlight} Times of minimum light.}
\tablecolumns{5}

\tablehead{
\colhead{HJD\_UTC} & \colhead{Uncertainty} & \colhead{Cycle} & \colhead{Eclipse} & \colhead{Instrument} \\
\colhead{(days)} & \colhead{(s)} & &
}

\startdata
$2455577.272644$ &$ 7$ &$7$ &Primary   &Hankasalmi \\
$2455355.627938$ &$41$ &$1$ &Secondary &Clay \\
$2455478.725045$ &$42$ &$4$ &Secondary &Clay \\
$2455519.757412$ &$31$ &$5$ &Secondary &Clay \\
$2455560.789627$ &$11$ &$6$ &Secondary &Clay \\
$2455519.757351$ &$16$ &$5$ &Secondary &MEarth \\
$2455683.887000$ &$20$ &$9$ &Secondary &MEarth \\
\enddata

\tablecomments{Estimated using the method of
  \citet{1956BAN....12..327K}, over an interval of $\pm 10^{-3}$ in
  normalized orbital phase.  These times of minimum are not used in
  our analysis (we fit for the ephemeris directly from the data) and
  are reported only for completeness.}

\end{deluxetable}

\subsection{Primary eclipse}

Using the timing of the single visit within the primary eclipse of UT
2010 April 3 from MEarth (see \S \ref{obs_detect_sect}) combined with
modeling of the secondary eclipses (\S \ref{obs_sec_sect}) and radial
velocities (\S \ref{spec_sect}) it was possible to predict the
primary eclipse times with reasonable accuracy.  During 2010-2011
full eclipses were observable at reasonable airmass ($< 2.0$) from
eastern Europe, Scandinavia, and western Asia, with the high
declination of the target favoring northerly latitudes during winter
to maximize dark hours at reasonable airmass.

The primary eclipse was observed using the $40\ {\rm cm}$ telescope at
Hankasalmi Observatory, Finland on UT 2011 January 15, starting
approximately 10 minutes before first contact, and continuing until
well after last contact.  Data were taken using a similar
$\ibess$ filter to the secondary eclipse observations (see \S
\ref{obs_sec_sect}) with a front-illuminated Kodak KAF-1001E CCD
packaged by Santa Barbara Instrument Group (SBIG; model STL-1001E).
This system response should be very similar to the one used for the
secondary eclipses, but nonetheless, any bandpass mismatch is a
potential concern for the accuracy of the photometric solution when
combining data taken with different instruments, and will be discussed
in more detail in \S \ref{mod_band_sect}.

Exposure times of $60\ {\rm s}$ were used, yielding a cadence of
approximately $2$ minutes, and the target was re-centered every $10$
exposures to keep the drift below $\approx 50$ pixels.  Data were
reduced using 2-D bias frames, and master dark and flat-field frames.
Light curves were then derived using the same methods and software as
for the other data-sets.

\subsection{Out of eclipse and additional secondary eclipses}
\label{obs_ooe_sect}

Following the discovery of atmospheric water vapor induced systematic
effects in the MEarth photometry (see \citealt{2011ApJ...727...56I}),
a customized $I$-band filter was procured for the start of the
2010-2011 season, with a bandpass designed to eliminate the strong
telluric water vapor absorption at wavelengths $>900\ {\rm nm}$, by
using an $895\ {\rm nm}$ ($50\%$ transmittance point) interference
cutoff on the same RG715 glass substrate as the original filter (where
the long-pass glass filter defines the blue end of the bandpass).
This was predicted to reduce the effect to largely insignificant
levels, and the resulting system response approximates the Cousins $I$
bandpass (as tabulated by \citealt{1990PASP..102.1181B}; see later in
this section).

Observations of \theobject\ using this new filter commenced on UT 2010
October 29, and continued until UT 2011 May 18.  Exposures of $2
\times 97\ {\rm s}$ were used at $20\ {\rm minute}$ cadence for times
out of eclipse.  $5283$ useful data points were obtained at airmasses
$< 2$, after discarding $400$ bad frames (pointing errors, severe
light loss due to clouds, and large scatter in the zeropoint solution
for the differential photometry).

Partial secondary eclipses were obtained with the same MEarth
telescope on UT 2010 November 19, UT 2011 February 9, and UT 2011 May
2.  Due to known systematic effects caused by persistence in the
MEarth CCDs, leading to offsets in magnitude depending on the cadence
(see \citealt{2011ApJ...736...12B}), and to remove any residual from
the out-of-eclipse variations, we allow these high-cadence
observations to have different photometric zero-points than the out of
eclipse data (and each other) in the final light curve analysis.

The out-of-eclipse data (excluding the high-cadence eclipse
observations) were searched for periodic variations using the method
outlined in \citet{2011ApJ...727...56I}.  Correction for the ``common
mode'' effect was still performed, as we discovered during the season
that replacing the filters had in fact increased the size (and changed
the sign) of the systematic effect rather than eliminating it.  This
problem is still under investigation, but we suspect it is due to
humidity and/or temperature dependence of the filter bandpass, an
effect also seen in the original Sloan Digital Sky Survey
(SDSS) Photometric Telescope filters \citep{2010AJ....139.1628D}.

\begin{figure}
\centering
\includegraphics[angle=270,width=3.1in]{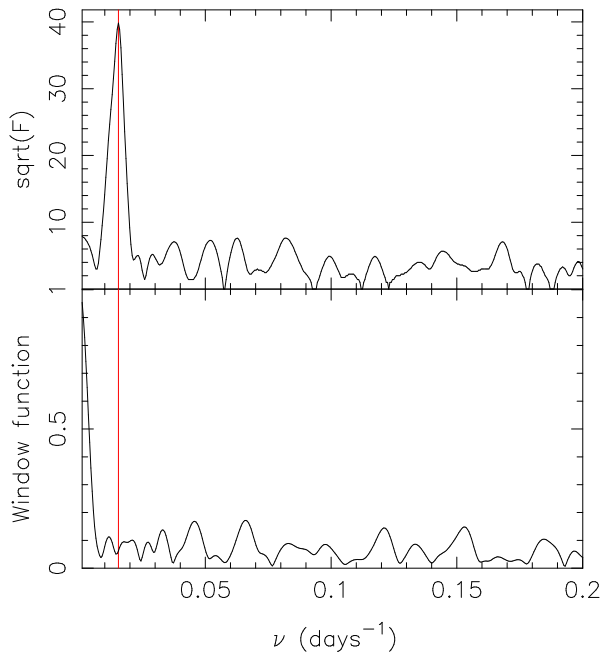}
\caption{``F-test periodogram'' for the MEarth 2010-2011 season out of 
  eclipse observations, plotting the square root of the F-test
  statistic comparing the null hypothesis of no variation to the
  alternate hypothesis of a sinusoidal modulation, as a function of
  frequency (top) and the corresponding window function (bottom).  The
  vertical line indicates the best-fitting period of $\theooeperiodsf\ \days$.}
\label{ooe_periodogram}
\end{figure}

After correcting for the ``common mode'' effect, the MEarth data show
a strong near-sinusoidal modulation with a $\theooeperiod \ \oneday$
period (see Figure \ref{ooe_periodogram}; the data are shown in \S
\ref{model_sect}) and a peak-to-peak amplitude of approximately $0.02\
{\rm mag}$.  Marginal evidence for a similar periodicity is found in
the 2009-2010 data, although the phase is not consistent and the
amplitude appears to be much smaller.  It is unclear if the amplitude
change results from the differing filter systems, but the change in
phase (if real) may be indicative of evolution of the photospheric
spot patterns giving rise to the modulations.

An improvement resulting from the new MEarth filter is a greatly
reduced color term when standardizing the photometry onto the Cousins
system.  The scatter in this relation also appears to be significantly
lower, which may also result in part from improvements made to our
flat-fielding strategy during the 2010-2011 season.  The MEarth
observing software automatically obtains measurements of equatorial
photometric standard star fields from the catalog of
\citet{1992AJ....104..340L} every night, and these are used to derive
photometric zero-points and calibrated photometry.  The typical
scatter in these zero-point solutions on a photometric night is $0.02\
{\rm mag}$.  We find a typical atmospheric extinction of $0.05\ {\rm
  mag/airmass}$ from MEarth data, which has been assumed henceforth.

A significant issue with the equatorial standard fields is they
contain very few red stars suitable for defining the red end of the
photometric response, and so color equations derived only from these
equatorial fields may be significantly in error when attempting to
standardize M-dwarf photometry.  Thankfully, MEarth serendipitously
targeted a number of M-dwarfs with photometry available on the Cousins
system from \citet{1990A&AS...83..357B}.  We selected all such objects
in the 2010-2011 MEarth database with data available on photometric
nights for calibration.  Two objects with clearly unreliable MEarth
data were discarded, leaving $8$ objects.  The following color
equation was derived from these stars, which span $\icous-K_{\rm 2MASS}$
colors of $2.2-4.2$:
\begin{eqnarray*}
\icous = \imear &-& (0.085 \pm 0.019) \\
                &+& (0.0204 \pm 0.0064)\ (\imear-K_{\rm 2MASS})
\end{eqnarray*}
In the case of \theobject, we estimate $\icous = 12.14 \pm 0.05$ using
this relation, where we have inflated the uncertainty to account for
variability and any errors introduced by the varying MEarth bandpass
as discussed above.  The observed $\icous-J$, $\icous-H$ and
$\icous-K$ colors are later used to infer effective temperatures and
luminosities for the members of the binary, and thus the distance to
the system.

\subsection{Literature data: proper motion and photometry}
\label{obs_lit_sect}

In Table \ref{photparams}, we summarize the photometric and
astrometric system properties gathered from literature data and
our photometry.  This star lies within the area covered by the Sloan
Extension for Galactic Understanding and Evolution (SEGUE) stripe 1260
photometry, so we retrieved the PSF magnitudes from Data Release 7
\citep{2009ApJS..182..543A}.  This object is flagged as saturated in
$i$-band (the image also shows a very unusual PSF and may be
corrupted), and the $u$-band photometry of M-dwarfs in SDSS is
corrupted by a well-documented red leak problem, so we omit these
filters.

\begin{deluxetable}{lr}
\tabletypesize{\normalsize}
\tablecaption{\label{photparams} Summary of the photometric and
  astrometric properties of the \theobject\ system.}
\tablecolumns{2}

\tablehead{
\colhead{Parameter} & \colhead{Value}
}

\startdata
$\alpha_{2000}$\tablenotemark{a,b}    & $11^h12^m42^s.35$ \\
$\delta_{2000}$\tablenotemark{a,b}    & $+76^\circ26\arcmin56\farcs3$ \\
$\mu_\alpha \cos \delta$\tablenotemark{b}       & $ 0\farcs131\ {\rm yr^{-1}}$ \\
$\mu_\delta$\tablenotemark{b}         & $-0\farcs108\ {\rm yr^{-1}}$ \\
\\
$g$\tablenotemark{c}                  &$15.618 \pm 0.010$ \\
$r$\tablenotemark{c}                  &$14.195 \pm 0.012$ \\
$z$\tablenotemark{c}                  &$12.027 \pm 0.012$ \\
\\
$I_{\rm C}$\tablenotemark{d}      &$12.14 \pm 0.05$ \\
\\
$J_{\rm 2MASS}$\tablenotemark{e}  & $10.591 \pm 0.023$ \\
$H_{\rm 2MASS}$\tablenotemark{e}  & $9.989 \pm 0.021$ \\
$K_{\rm 2MASS}$\tablenotemark{e}  & $9.728 \pm 0.017$ \\
\multicolumn{2}{p{2.8in}}{}  
\enddata

\tablenotetext{a}{Equinox J2000.0, epoch 2000.0.}
\tablenotetext{b}{From \citet{2005AJ....129.1483L}.}
\tablenotetext{c}{From SDSS Data Release 7.  Note that these
  uncertainties do not account for variability, and are therefore
  underestimates.}
\tablenotetext{d}{From \S \ref{obs_ooe_sect}.}
\tablenotetext{e}{We quote the combined uncertainties from the 2MASS
  catalog, noting that the intrinsic variability of our target means
  in practice that these are underestimates.}

\end{deluxetable}

\begin{figure}
\centering
\includegraphics[angle=0,width=3.2in]{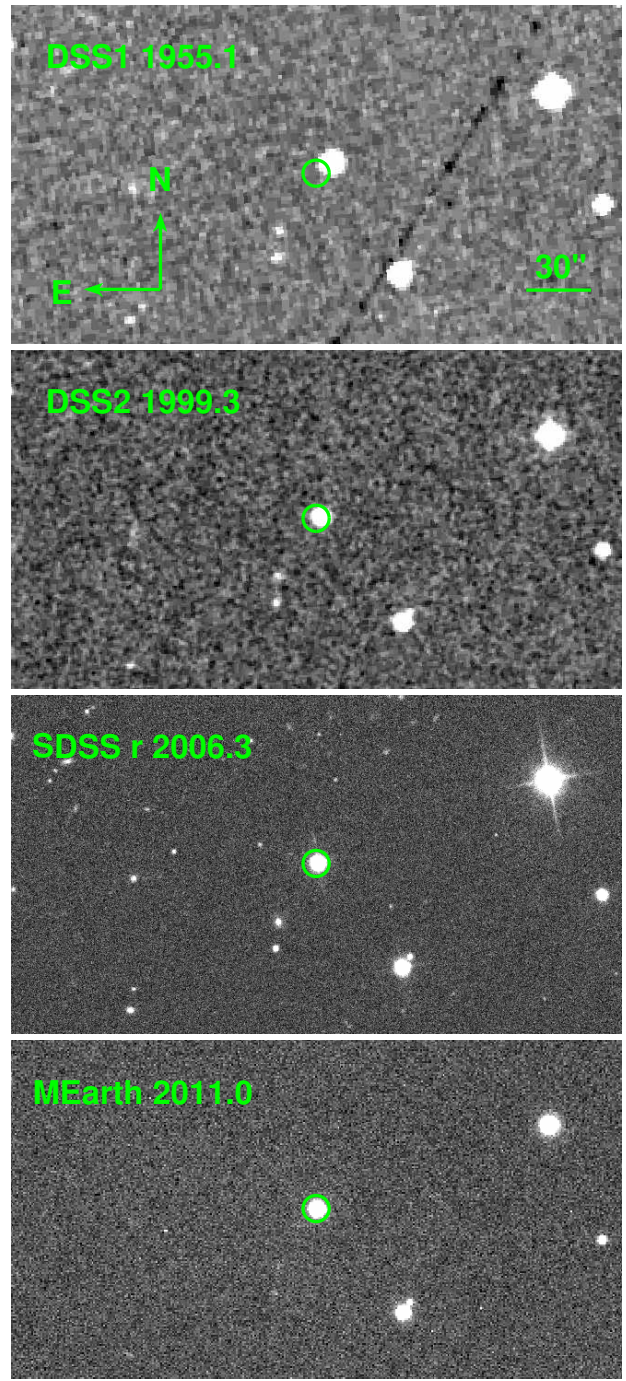}
\caption{Images of \theobject\ centered on the position as
  measured from the MEarth data.  The circle shows the approximate
  position and size of the $8\arcsec$ (radius) photometric aperture
  used to derive the MEarth light curves.  Data are from
  the first and second epoch Palomar sky surveys as provided by the
  Digitized Sky Survey (DSS; top and center panels), SDSS, and the
  MEarth master image (bottom panel).  The approximate epochs are
  shown on the images, and all four panels have the same center,
  scale and alignment on-sky.}
\label{third_light_images}
\end{figure}

The moderately high proper motion of our target allows the
contribution of any background objects to the system light to be
constrained using previous epochs of imaging.  In Figure
\ref{third_light_images}, we show imaging data with the position and
size of the MEarth photometric aperture (which is representative of 
the aperture sizes used for all the photometry) overlaid.  The DSS1
image indicates the aperture is unlikely to contain any significant
flux from background objects, and the SDSS image (which has the
highest angular resolution) does not appear elongated.

\subsection{Intermediate-resolution optical spectroscopy: spectral
  typing and metallicity constraints}
\label{obs_spec_sect}

\begin{figure}
\centering
\includegraphics[angle=270,width=3.2in]{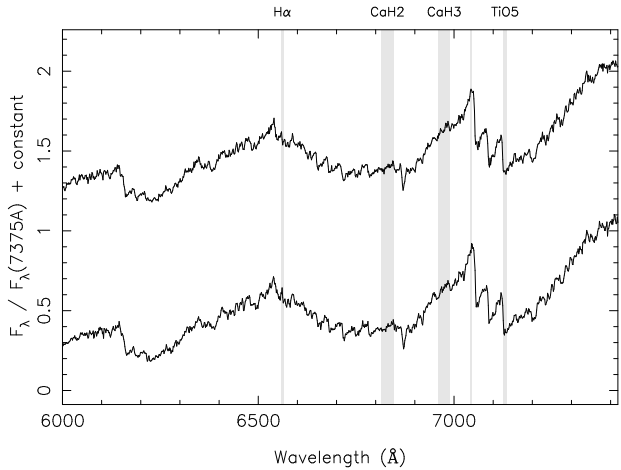}
\caption{Intermediate-resolution FAST spectra of \theobject\ showing
  the TiO and CaH molecular bands typically used for M-dwarf spectral
  classification.  Two exposures were obtained and are shown offset in
  the vertical direction for clarity.  Overlaid in gray are the
  locations of the H$\alpha$ line, and the CaH2, CaH3 and TiO5 band
  indices defined by \citet{1995AJ....110.1838R}.}
\label{spectrum}
\end{figure}

A single epoch intermediate-resolution ($R \simeq 1200$) spectrum of our
target was obtained using the FLWO $1.5\ {\rm m}$ Tillinghast
reflector and FAST spectrograph \citep{1998PASP..110...79F} on UT 2011
March 16.  The default setting of a $300\ {\rm g/mm}$ grating and
$3\arcsec$ slit was used to obtain two exposures of $300\ {\rm s}$.
The spectra were reduced using standard procedures for long slit
spectra in IRAF\footnote{IRAF is distributed by the National Optical 
Astronomy Observatories, which are operated by the Association of
Universities for Research in Astronomy, Inc., under cooperative
agreement with the National Science Foundation.}
\citep{1993ASPC...52..173T}, using a HeNeAr arc exposure taken
inbetween the two target exposures to ensure accurate wavelength
calibration.  The 2-D spectra were divided by a flat field before
extraction to remove the majority of the fringing seen at the reddest
wavelengths.  Relative flux calibration was performed using the
spectrophotometric standard BD+26~2606.  We show the resulting spectra
in Figure \ref{spectrum}.

Using the TiO5 band index defined by \citet{1995AJ....110.1838R}, we
find a composite spectral type of M3.7 (note that the uncertainty in
this value is at least $0.5$ subclass), and using {\sc The Hammer}
\citep{2007AJ....134.2398C}, we find M3-M4 (visual inspection
indicates the observed spectrum is closer to M4).  The spectral type
and the measured colors are in good agreement using the empirical
colors of Galactic disk M-dwarfs from \citet{1992ApJS...82..351L} for
the Johnson/Cousins/CIT system and \citet{2005PASP..117..706W} for the
SDSS/2MASS.

It is also clear from Figure \ref{spectrum} that no H$\alpha$ line is
seen in the spectrum.  Such behavior is typical of inactive field
dwarfs at these spectral types when observed at low resolution, where
the equivalent width of the H$\alpha$ absorption line becomes close to
zero at M4-M5 (e.g. \citealt{2002AJ....123.3356G}; see their Figure 5,
noting that the TiO5 index was $0.42$ for \theobject).  The presence
of H$\alpha$ emission in M-dwarf spectra is usually indicative of high
activity levels.

The relative strengths of the TiO and CaH bands in optical spectra
have been used to identify and classify M-subdwarfs
\citep{1997AJ....113..806G}, and more generally proposed as a
method for estimating metallicities of M-dwarfs
(e.g. \citealt{2006PASP..118..218W}).  While detailed calibrations
(e.g. using band indices) do not yet appear to be available, we
performed a visual comparison between Figure \ref{spectrum} and
the spectra shown in \citet{1997AJ....113..806G}.  The strong TiO
bands seen in our target indicate it is not extremely metal poor.  We
therefore estimate it has metallicity $[{\rm Fe/H}] > -1$, and is
probably closer to solar (or greater).

\subsection{High-resolution optical spectroscopy: radial velocities}
\label{obs_rv_sect}

High-resolution spectroscopic observations were obtained using the
TRES fiber-fed \'echelle spectrograph on the FLWO $1.5\ {\rm m}$
Tillinghast reflector.  We used the medium fiber ($2\farcs3$ projected
diameter), yielding a resolving power of $R \simeq 44\,000$.

Observations commenced on UT 2010 May 1, and continued until UT 2011
April 13, with $43$ epochs acquired.  Total exposure times ranged from
45 to 80 minutes, broken into sequences of 3 or 4 individual exposures.
The spectra were extracted using a custom built pipeline designed to
provide precise radial velocities for \'echelle spectrographs.  The
procedures are described in more detail in
\citet{2010ApJ...720.1118B}, and will be summarized briefly below.

In order to effectively remove cosmic rays, the sets of raw images
were median combined.  We used a flat field to trace the \'echelle
orders and to correct the pixel to pixel variations in CCD response
and fringing in the red-most orders, then extracted one-dimensional
spectra using the ``optimal extraction'' algorithm of
\citet{1985MNRAS.213..971H} (see also \citealt{1986PASP...98..609H}).
The scattered light in the two-dimensional raw image was determined
and removed by masking out the signal in the \'echelle orders and
fitting the inter-order light with a two-dimensional polynomial.

Thorium-Argon (ThAr) calibration images were obtained through the
science fiber before and after each stellar observation.  The two
calibration images were combined to form the basis for the fiducial
wavelength calibration.  TRES is not a vacuum spectrograph and
is only temperature controlled to $0.1 {\rm ^\circ C}$.  Consequently, the
radial velocity errors are dominated by shifts due to pressure,
humidity and temperature variations.  In order to successfully remove
these large variations ($> 1.5\ \kms$), it is critical that the
ThAr light travels along the same optical path as the stellar light
and thus acts as an effective proxy to remove these variations.  We have
therefore chosen to sandwich the stellar exposure with two ThAr frames
instead of using the simultaneous ThAr fiber, which may not exactly
describe the induced shifts in the science fiber and can also lead to
bleeding of ThAr light into the science spectrum from the strong argon
lines, especially at redder wavelengths.  The pairs of ThAr exposures
were co-added to improve the signal to noise ratio, and centers of the
ThAr lines found by fitting a Gaussian function to the line profiles.
A two-dimensional fifth order Legendre polynomial was used to
describe the wavelength solution.

\section{Spectroscopic analysis}
\label{spec_sect}

Radial velocities were obtained using the two-dimensional
cross-correlation algorithm {\sc todcor} \citep{1994ApJ...420..806Z},
which uses templates matched to each component of a spectroscopic
binary to simultaneously derive the velocities of both stars, and an
estimate of their light ratio in the spectral bandpass.

We used a single epoch observation of Barnard's star
(\citealt{1916AJ.....29..181B}; also known as Gl~699) taken on UT 2008
October 20 as the template for the {\sc todcor} analysis.  Barnard's
star has a spectral type of M4 \citep{1991ApJS...77..417K}, which
should be a reasonable match to both components.  Correlations were
performed using a wavelength range of 7063 to 7201{\AA} in order 41 of
the spectrum to derive the velocities, since this region contains
strong molecular features (mostly from TiO), which are rich in radial
velocity information.  We assume a barycentric radial velocity of
$-110.3 \pm 0.5\ \kms$ for Barnard's star (where the stated
uncertainty reflects our estimate of the systematic errors), derived 
from presently unpublished CfA Digital Speedometer measurements
spanning $17\ {\rm years}$.

To derive the light ratio, we ran {\sc todcor} using a range of fixed
values of this quantity, seeking the maximum correlation over all
epochs.  In this way, consistency between the different epochs is
enforced, in accordance with the expectation that the true value of
this quantity should be (approximately) constant.  This procedure
improves the stability of the solution and minimizes systematic errors
in the velocities.  We find a best-fitting light ratio of $L_2/L_1 =
0.528 \pm 0.05$, where we adopt a conservative estimate of the
uncertainty.

The radial velocities derived from our analysis are reported in Table
\ref{rv-data}.  We omit four velocities with peak correlation values
below $C_{\rm peak} = 0.5$ from the table and the remainder of the
analysis as the low correlation values indicate these velocities are
likely to be unreliable or to exhibit large scatter.  The uncertainty
in the radial velocities (estimated from our fitting procedure; see \S
\ref{model_sect}) is $0.26\ \kms$ for the primary, and $0.58\ \kms$
for the secondary, with an additional systematic uncertainty,
estimated to be $0.5\ \kms$ (see above), in the velocity zero-point
system.

\begin{deluxetable}{lrrrr}
\tabletypesize{\normalsize}
\tablecaption{\label{rv-data} Radial velocity data.}
\tablecolumns{5}

\tablehead{
\colhead{HJD\_UTC\tablenotemark{a}} & \colhead{Phase\tablenotemark{b}}
& \colhead{$v_1$\tablenotemark{c}} & \colhead{$v_2$\tablenotemark{c}}
& \colhead{$C_{\rm peak}$\tablenotemark{d}} \\ 
\colhead{(days)} & & \colhead{($\kms$)} & \colhead{($\kms$)} &
}

\startdata
$2455317.6481$ &$0.67269$ &$-49.45$ &$-78.49$ &$0.560$ \\
$2455320.6671$ &$0.74626$ &$-41.11$ &$-90.08$ &$0.635$ \\
$2455321.6714$ &$0.77074$ &$-38.58$ &$-93.77$ &$0.643$ \\
$2455341.6568$ &$1.25780$ &$-79.89$ &$-33.62$ &$0.586$ \\
$2455342.6784$ &$1.28270$ &$-79.29$ &$-34.71$ &$0.506$ \\
$2455343.6545$ &$1.30649$ &$-78.44$ &$-36.09$ &$0.541$ \\
$2455344.6632$ &$1.33107$ &$-78.21$ &$-36.76$ &$0.577$ \\
$2455345.6848$ &$1.35597$ &$-76.63$ &$-38.66$ &$0.642$ \\
$2455346.6622$ &$1.37979$ &$-75.23$ &$-40.34$ &$0.536$ \\
$2455347.6488$ &$1.40383$ &$-74.26$ &$-42.12$ &$0.654$ \\
$2455348.6433$ &$1.42807$ &$-72.38$ &$-44.19$ &$0.602$ \\
$2455366.6561$ &$1.86706$ &$-35.28$ &$-97.31$ &$0.627$ \\
$2455367.6852$ &$1.89214$ &$-37.23$ &$-96.02$ &$0.588$ \\
$2455368.6748$ &$1.91626$ &$-40.62$ &$-90.98$ &$0.670$ \\
$2455369.6903$ &$1.94101$ &$-44.54$ &$-84.52$ &$0.576$ \\
$2455373.6462$ &$2.03742$ &$-65.72$ &$-53.84$ &$0.547$ \\
$2455374.6460$ &$2.06178$ &$-69.12$ &$-48.82$ &$0.662$ \\
$2455375.6455$ &$2.08614$ &$-72.87$ &$-43.53$ &$0.582$ \\
$2455581.9638$ &$7.11433$ &$-75.90$ &$-39.72$ &$0.750$ \\
$2455582.9270$ &$7.13780$ &$-77.72$ &$-37.42$ &$0.788$ \\
$2455583.8762$ &$7.16093$ &$-79.01$ &$-35.40$ &$0.811$ \\
$2455584.9160$ &$7.18627$ &$-79.99$ &$-34.08$ &$0.806$ \\
$2455586.9587$ &$7.23606$ &$-80.51$ &$-33.75$ &$0.854$ \\
$2455587.9329$ &$7.25980$ &$-80.20$ &$-33.98$ &$0.829$ \\
$2455605.9077$ &$7.69786$ &$-46.45$ &$-82.16$ &$0.816$ \\
$2455606.9374$ &$7.72296$ &$-43.79$ &$-87.07$ &$0.624$ \\
$2455607.9274$ &$7.74709$ &$-41.08$ &$-90.00$ &$0.797$ \\
$2455608.8926$ &$7.77061$ &$-38.34$ &$-93.22$ &$0.757$ \\
$2455611.8887$ &$7.84363$ &$-34.67$ &$-99.11$ &$0.731$ \\
$2455614.8719$ &$7.91633$ &$-40.32$ &$-91.05$ &$0.741$ \\
$2455615.8710$ &$7.94068$ &$-44.85$ &$-84.66$ &$0.822$ \\
$2455616.9096$ &$7.96599$ &$-50.16$ &$-76.85$ &$0.742$ \\
$2455617.8577$ &$7.98910$ &$-55.18$ &$-69.76$ &$0.788$ \\
$2455650.7940$ &$8.79179$ &$-36.93$ &$-95.63$ &$0.780$ \\
$2455651.7573$ &$8.81526$ &$-35.56$ &$-97.88$ &$0.732$ \\
$2455652.8008$ &$8.84070$ &$-34.89$ &$-98.32$ &$0.808$ \\
$2455656.7773$ &$8.93761$ &$-44.06$ &$-85.15$ &$0.814$ \\
$2455659.7417$ &$9.00985$ &$-59.54$ &$-62.78$ &$0.815$ \\
$2455664.7613$ &$9.13218$ &$-77.22$ &$-38.03$ &$0.826$ \\
\enddata

\tablenotetext{a}{Heliocentric Julian Date of mid-exposure, in the UTC
time-system.}
\tablenotetext{b}{Normalized orbital phase.}
\tablenotetext{c}{Barycentric radial velocity of stars 1 and 2.}
\tablenotetext{d}{Peak normalized cross-correlation from {\sc
    todcor}.  These are later used for weighting the radial velocity
  points in the solution.}

\end{deluxetable}

We find no evidence for ``peak pulling'' effects close to the systemic
velocity, which might indicate the need for additional rotational
broadening of the template to match the target.  Barnard's star is a
very slow rotator (e.g. \citealt{1998AJ....116..429B}), so this and
the high correlation values obtained without need for additional
rotational broadening indicate both components of our target have low
$v \sin i$, as expected.

\section{Model}
\label{model_sect}

For EB systems showing moderate or high eccentricities, the commonly
exploited property of being able to separate the photometric and
spectroscopic solutions no longer holds as the eccentricity affects
both.  Typically, the orbital period $P$, epoch of primary eclipse
$T_0$, and $e \cos \omega$ (where $e$ is eccentricity and $\omega$ is
the argument of periastron) are better determined by the photometry
(by eclipse timing; the latter is related to the departure of the
secondary eclipses in phase from being exactly halfway between the
primary eclipses, although note that the commonly-used one-to-one
relation between these quantities is approximate), and the $e \sin
\omega$ component is usually best constrained by the velocities (but
also affects the relative durations of the two eclipses, and in
general eccentricity also influences their shape). Therefore, in order
to leverage the best possible constraints on the system orbit, we
adopt a joint solution of all the light curves and the radial
velocities simultaneously.  This also greatly simplifies the error
analysis.

We proceed by first discussing the basic model and assumptions in \S 
\ref{mod_assum_sect}, the method of solution in \S \ref{mod_sol_sect},
and then the effects of four potentially important sources of
systematic error in the system parameters we derive: spots (\S
\ref{mod_spot_sect}), bandpass mismatch (\S \ref{mod_band_sect}),
limb darkening (\S \ref{mod_ld_sect}), and third light (\S
\ref{mod_l3_sect}).  In \S \ref{mod_sum_sect}, we summarize our best
estimates of the system parameters and their corresponding
uncertainties.

\subsection{Basic model and assumptions}
\label{mod_assum_sect}

To model the system, we used the light curve generator from the
popular {\sc jktebop} code
\citep{2004MNRAS.351.1277S,2004MNRAS.355..986S}, which is in turn
based on the eclipsing binary orbit program ({\sc ebop};
\citealt{1981AJ.....86..102P}).  This model approximates the stars
using biaxial ellipsoids \citep{1972ApJ...174..617N}, which is
appropriate for well-detached systems such as the present one.  We
modified the model to perform the computations in double precision, to
account for light travel time, which is necessary due to the large
semimajor axis, and to incorporate a simple prescription for the
effect of photospheric spots (discussed in \S \ref{mod_spot_sect}).
The light curve model was coupled with a standard Keplerian orbital
model for the radial velocities (also including the light travel
time).  We correct the orbital period from the heliocentric frame to
the system barycenter \citep{1976Obs....96..153G} when computing the
physical parameters, following the usual convention of reporting the
orbital period $P$ as measured in the heliocentric frame (i.e. without
the correction) in the tables.

We fit all of the Clay, Hankasalmi, and 2010-2011 season MEarth data,
the radial velocities, and the spectroscopic light ratio
simultaneously.  We also include the single primary eclipse of UT 2010
April 3 to improve the constraint on the ephemeris, assuming the
bandpass was approximately the same as the $\ibess$ filter (this makes
little difference in practice due to the sparse coverage).  The
remainder of the MEarth 2009-2010 season data were omitted as these
are of quite poor quality and were taken in a different bandpass from
the remainder, so do not usefully constrain the model after accounting
for the extra free parameters which must be added in order to fit them.

Table \ref{model_pars} summarizes the assumptions made in our light
curve model.  The light curve parameter set generally follows {\sc
jktebop}, and has been chosen to minimize correlations between
parameters where possible.  We add a term $k_l (X - 1)$ to the model
magnitudes, where $X$ is airmass, and $l$ is an index for the
different light curves, to account for differential atmospheric
extinction between the target and comparison stars, presumably due to
color-dependence.  A different $k_l$ value was allowed for each light
curve $l$.  This has only been used for fitting the Clay secondary
eclipse curves and 2010-2011 MEarth data (see \S \ref{obs_sec_sect}),
where it seems to be necessary to include this term to obtain an
acceptable fit to the out of eclipse portions.  We are unsure as to
the exact origin of the baseline curvature in the Clay data, and note
that the choice of a linear function of airmass is arbitrary and may
not be correct.  This is further evidenced by the inconsistent sign
and magnitude of this term in Table \ref{fit-spot-params}.  We suggest
the best approach to rectifying this problem would be to obtain higher
quality secondary eclipse observations.

\begin{deluxetable*}{lrll}
\tabletypesize{\normalsize}
\tablecaption{\label{model_pars} Parameters and priors used in the basic model.}
\tablecolumns{4}

\tablehead{
\colhead{Parameter} & \colhead{Value} & \colhead{Prior\tablenotemark{a}} & \colhead{Description}
}

\startdata
$J_{\rm MEarth}$  &varied &uniform       &Central surface brightness ratio (secondary/primary) in $\imear$ \\
$J_{\rm Bessell}$ &varied &uniform       &Central surface brightness ratio (secondary/primary) in $\ibess$ \\
$(R_1+R_2)/a$   &varied &uniform         &Total radius divided by semimajor axis \\
$R_2/R_1$       &varied &uniform         &Radius ratio \\
$\cos i$        &varied &uniform (isotropic)        &Cosine of orbital inclination \\
$e \cos \omega$ &varied &uniform in $e$,$\omega$  &Eccentricity times cosine of argument of periastron \\
$e \sin \omega$ &varied &uniform in $e$,$\omega$  &Eccentricity times sine of argument of periastron \\
\\
$L_2/L_1$       &$0.528 \pm 0.05$        &Gaussian  &$\imear$-band light ratio \\
\\
$u_1$           &$-0.0765$               &uniform   &Primary limb darkening linear coefficient\tablenotemark{b} \\
$u'_1$          &$1.0369$                &uniform   &Primary limb darkening sqrt coefficient\tablenotemark{b} \\
$u_2$           &$0.0561$                &uniform   &Secondary limb darkening linear coefficient\tablenotemark{b} \\
$u'_2$          &$0.9335$                &uniform   &Secondary limb darkening sqrt coefficient\tablenotemark{b} \\
$\beta_1$       &$0.32$ &\ldots    &Primary gravity darkening exponent\tablenotemark{c} \\
$\beta_2$       &$0.32$ &\ldots    &Secondary gravity darkening exponent\tablenotemark{c} \\
$\kappa_1$      &computed &\ldots  &Primary reflection effect coefficient \\
$\kappa_2$      &computed &\ldots  &Secondary reflection effect coefficient \\
$L_3$           &$0$      &uniform  &Third light divided by total light\tablenotemark{b} \\
$\Phi$          &$0$      &\ldots  &Tidal lead/lag angle \\
$\delta$        &$2^\circ$ &\ldots &Integration ring size\tablenotemark{d} \\
\\
$F_1$           &varied\tablenotemark{e} &uniform   &Primary rotation parameter \\
$a_1$           &varied\tablenotemark{e} &uniform   &Primary out-of-eclipse sine coefficient \\
$b_1$           &varied\tablenotemark{e} &uniform   &Primary out-of-eclipse cosine coefficient \\
$F_2$           &varied\tablenotemark{e} &uniform   &Secondary rotation parameter \\
$a_2$           &varied\tablenotemark{e} &uniform   &Secondary out-of-eclipse sine coefficient \\
$b_2$           &varied\tablenotemark{e} &uniform   &Secondary out-of-eclipse cosine coefficient \\
\\
$q$             &varied &uniform   &Mass ratio \\
$K_1+K_2$       &varied &modified Jeffreys  &Sum of radial velocity
semiamplitudes  \\
                &       &$K_a = 0.02\ \kms$ &Prior $P(K_1+K_2) \propto 1/(K_1+K_2+K_a)$ \\
$\gamma$        &varied &uniform   &Systemic radial velocity \\
\\
$P$             &varied &uniform   &Orbital period (heliocentric) \\
$T_0$           &varied &uniform   &Epoch of primary conjunction (HJD\_UTC) \\
$T_{\rm sec}$   &computed &\ldots  &Epoch of secondary conjunction (HJD\_UTC) \\
\\
$z_l$           &varied &uniform   &Baseline magnitude for light curve $l$\tablenotemark{f} \\
$s_l$           &varied &Jeffreys  &Error scaling factor for light curve $l$\tablenotemark{f} \\
$k_l$           &varied &uniform   &Airmass coefficient for light curve $l$ (where used)\tablenotemark{f} \\
$C_l$           &varied &uniform   &Common mode coefficient for light curve $l$ (where used)\tablenotemark{f} \\
\\
$s_r$           &varied &Jeffreys  &Error (if $C_{\rm peak}=1$) for radial velocity curve $r$\tablenotemark{f} \\
\enddata

\tablenotetext{a}{See \S \ref{mod_sol_sect} for discussion.}
\tablenotetext{b}{Varied in later sections.}
\tablenotetext{c}{Values appropriate for convective atmospheres.}
\tablenotetext{d}{Our tests indicate this value results in a maximum
  error of $3 \times 10^{-5}\ {\rm mag}$ in the eclipse depth
  (compared to a model with $0.1^\circ$), which should be sufficient
  for our photometry.}
\tablenotetext{e}{Parameters varied only for one of the stars in each
  simulation.}
\tablenotetext{f}{$l$ and $r$ are indices for the different light
  curves and radial velocities for each component of the system.}

\end{deluxetable*}

Gravity darkening coefficients were fixed at values appropriate for
convective atmospheres \citep{1967ZA.....65...89L}, although gravity
darkening is unimportant with the precision of the present light
curves at such low rotation rates.  The reflection effect was computed
rather than fitting for it, although again the large semimajor axis
means it is unimportant.

We used a square root limb darkening law \citep{1992A&A...259..227D},
of the form:
\begin{equation}
{I_\lambda(\mu)\over{I_\lambda(1)}} = 1 - u (1 - \mu) - u' (1 - \sqrt{\mu})
\label{ld_law}
\end{equation}
where $I_\lambda(\mu)$ is the specific intensity, and $\mu = \cos
\theta$, where $\theta$ is the angle between the surface normal and
the line of sight.  As discussed by \citet{1993AJ....106.2096V}, the
square root law is a better approximation to the specific intensity
distribution given by model atmospheres of late-type stars in the NIR
than the other common two-parameter laws (quadratic and logarithmic)
implemented in {\sc jktebop}.  We also verified this by comparing the
\citet{2000A&A...363.1081C} four-parameter law (which we assumed to
be the best representation of the original {\sc phoenix} model) with
the two-parameter laws for typical M-dwarf stellar parameters.

For our baseline model, we fixed the limb darkening coefficients to
values for the $\icous$ filter derived from {\sc phoenix} model
atmospheres by \citet{2000A&A...363.1081C}, using $\teff = 3130\ {\rm
  K}$ for the primary, and $\teff = 3010\ {\rm K}$ for the secondary,
$\log g = 5.0$ and solar metallicity.  We assume the same limb
darkening law applies to all the $I$ filters (the error introduced by
this assumption is examined in \S \ref{mod_ld_sect}).

No evidence was found for any background sources in the photometric
aperture (see \S \ref{obs_lit_sect}), and the radial velocity
observations do not show additional lines or velocity drifts that
might indicate a comoving, physically associated companion to the EB
system.  We therefore assume no third light ($L_3 = 0$), but see also
\S \ref{mod_l3_sect}.

\subsection{Method of solution}
\label{mod_sol_sect}

To derive parameters and error estimates, we adopt a variant of the
popular Markov Chain Monte Carlo (MCMC) analysis frequently applied in
cosmology and for analysis of exoplanet radial velocity and light
curve data (e.g. \citealt{2004PhRvD..69j3501T,2005AJ....129.1706F}).
We use Adaptive Metropolis \citep{haario2001} rather than the standard
Metropolis-Hastings method with the Gibbs sampler.  The adaptive
method allows adjustment of the proposal distributions while the chain
runs, and is thus simpler and faster to operate, obviating the need
for manual tuning of the proposal distributions typical of the more
conventional methods.  This method uses a multivariate Gaussian
proposal distribution to perturb all parameters simultaneously.

Compared to the simpler Monte Carlo and bootstrapping methods (e.g. as
implemented in {\sc jktebop}) these methods have an important
advantage for the present case where a heterogeneous set of light
curves and radial velocities must be analyzed: it is possible to
estimate the appropriate inflation of the observational uncertainties
self-consistently and simultaneously with the other model parameters
(e.g. \citealt{2005ApJ...631.1198G}).  This allows the correct
relative weighting of the various observational data-sets to be
decided essentially by their residuals from the best-fitting model,
and the uncertainty in this weighting to be propagated to the final
parameters.  Of course, this assumes the correct model has been chosen
for the data, so these methods must be used carefully.

In the Monte Carlo simulations, we used the standard
Metropolis-Hastings acceptance criterion, accepting the new point with
probability:
$$
P_{\rm acc} = \left\{ \begin{array}{r}
1, P(M_{i+1}|D) \ge P(M_i|D) \\
{P(M_{i+1}|D)\over{P(M_i|D)}}, P(M_{i+1}|D) < P(M_i|D)
\end{array} \right.
$$
where $P(M_i|D)$ and $P(M_{i+1}|D)$ are the posterior probabilities of
the previous and current points, respectively, $M_i$ represents the
$i$th model, and $D$ the data.  The previous point in the chain was
repeated if the new point was not accepted (this is required to
correctly implement the method).  The appropriate ratio of posterior
probabilities (``Bayes factor'') was:
\begin{equation}
{P(M_{i+1}|D)\over{P(M_i|D)}} = {P(M_{i+1})\over{P(M_i)}}\ \exp(-\Delta \chi^2/2)
\label{bayes_factor}
\end{equation}
where the factors in the right-hand side are the ratio of prior
probabilities, and the usual (least squares) likelihood ratio,
respectively.

Priors assumed for each parameter are detailed in Table
\ref{model_pars}, and are chosen to be uninformative.  In most cases,
this choice is a uniform improper prior (labeled ``uniform'' in the
table).  For the orbital inclination, we adopt a uniform prior on
$\cos i$, which produces an isotropic distribution of orbit normals,
and for the eccentricity, we use a uniform prior on $e$ and $\omega$,
rather than on the jump parameters $e \cos \omega$ and $e \sin
\omega$ (the latter would produce a prior in $e,\omega$ proportional
to $e$; e.g. \citealt{2006ApJ...642..505F}).

As discussed by \citet{2005ApJ...631.1198G}, the appropriate choice of
uninformative prior for ``scale parameters'' such as radial velocity
amplitude or orbital period is a Jeffreys prior (or a modified
Jeffreys prior for parameters which can be zero).  This enforces scale
invariance (equal probability per decade).  In the present case, some
of these parameters (particularly the period) are so well-determined
the choice of prior is unimportant, so for simplicity we have only
used non-uniform priors on the less well-determined parameters where
the prior is important.

As discussed earlier, and in \citet{2005ApJ...631.1198G}, it is
possible to fit for the appropriate inflation of the observational
uncertainties simultaneously with the model parameters.  This has been
done via the $s_k$ parameters shown in Table \ref{model_pars}.  It is
conventional to apply this inflation by multiplication for light
curves, and for RV by adding an extra ``jitter'' contribution in
quadrature.  We follow these conventions here.  Note that additional
factors appear in the likelihood ratio of Eq. (\ref{bayes_factor})
when doing this, which have been subsumed into the $\Delta \chi^2$ in
the equation.

For the RV, it is difficult to estimate reliable uncertainties from
cross correlation functions, so instead we derive them during the
simulation.  We weight each point by $C_{\rm peak}^2$ (the square of
the peak normalized cross-correlation value; see Table \ref{rv-data}),
where we find $C_{\rm peak}$ is approximately proportional to the
signal to noise ratio in typical observations at similar signal to
noise and peak correlation values to those seen in this work.  This
procedure is essentially equivalent to photon-weighting, except the
cross-correlation accounts for all sources of uncertainty rather than
only photon noise.  The uncertainty corresponding to a peak
correlation of unity is derived during the simulation, and we allow
separate values of this quantity for each star, labeled $s_1$ for the
primary, and $s_2$ for the secondary, in the tables.  The resulting
per-data-point uncertainties used in the fitting are given by
$\sigma_{ij} = s_i / C_{\rm peak,j}$ for the $i$th star and $j$th
radial velocity point.

All Monte Carlo simulation results reproduced in this paper are
derived from chains of $2 \times 10^6$ points in length, where the
first $10\%$ of the points were used to initialize the parameter
covariance matrix for the Adaptive Metropolis method (starting from
the initial parameters and covariance matrix derived using a
Levenberg-Marquardt minimization\footnote{\tt
  http://www.ics.forth.gr/\~{}lourakis/levmar/}), and the next $40\%$
were used to ``burn in'' the chain.  This was found to be sufficient
to ensure the chains were very well converged, and all of these first
$50\%$ of the points were discarded, leaving the remainder ($10^6$
points) for parameter estimation.  Correlation lengths in all
parameters were $< 200$ points.

We report the median and $68.3\%$ central confidence intervals as the
central value and uncertainty for all parameters.  Reduced $\chi^2$
values for all the model fits were unity, by construction.

\subsection{Spots}
\label{mod_spot_sect}

In \theobject, the presence of spots is clearly indicated by the
observed out of eclipse modulation described in \S \ref{obs_ooe_sect}.

Spots present severe difficulty for the analysis of eclipsing binary
light curves because they cause systematic errors in the measurements
derived from the eclipses, principally their depths, which determine
the geometric parameters $J$, $(R_1+R_2)/a$ and $i$ (and by extension,
$R_2/R_1$).  Spots occulted during eclipse temporarily reduce the
depth as they are crossed, and spots not occulted during eclipse
increase it, because the surface brightness of the photosphere under
the eclipse chord is greater than would be inferred from the
out-of-eclipse baseline level.

The difficulty in solving for physical system parameters arises
because the true spot distribution is not usually known.
Out-of-eclipse modulations, and in non-synchronized systems,
modulation of the eclipses themselves, are sensitive only to the
longitudinal inhomogeneity in the spot distribution, except in special
cases where spots are crossed during eclipse and cause detectable
deviations from the usual light curve morphology (this can be 
difficult to detect in systems with radius ratios close to unity,
because the deviations then last a large fraction of the eclipse
duration).  Any longitudinally homogeneous component, such as a polar
spot viewed equator-on, or a homogeneous surface coverage of small
spots, cannot usually be detected from light curves.

This can cause undetected systematic errors in the radii and effective
temperatures derived from the light curves.
\citet{2010ApJ...718..502M} discuss this issue in some detail for the
best-measured literature EBs, finding this effect could produce up to
$6\%$ systematic errors in the radii for stars with spots of $30\%$
filling factor, when the spots are concentrated at the pole.

The influence of the spectroscopic light ratio is not commonly
discussed with regard to spots, but this is important because it 
provides complementary information to the photometry, on the
difference in spot coverage between the two stars.  For example,
consider the case of a large coverage of longitudinally homogeneous,
polar spots that are not crossed during eclipse.  If these spots are
distributed on both stars, the light ratio is unaffected but both
eclipses will appear deeper.  However, if the spots are on only one
star, the effect on the photometry is the same, but the light ratio is
altered because the spotted star appears darker.  This will produce a
discrepancy between the light ratio derived only from the light curve
parameters, and the spectroscopic value.  More generally, it is
possible to place constraints on the difference in the overall spot
coverage between the binary components, using the spectroscopic
information.

One unusual feature of the present system among low-mass eclipsing
binaries showing signs of spots is the spin and orbit appear to not be
synchronized.  Because of this, it is possible to measure eclipses at
different rotational phases, and thus different spottedness of the
visible stellar hemispheres.  While this does not necessarily resolve
the difficulty of determining the longitudinally homogeneous component
of the spot distribution, it does provide additional information on
the longitudinally inhomogeneous component, specifically which star
the spots are located on.  This information is not usually available
if only out of eclipse modulations are seen.  It is common to assume
the spots are on both binary components, which may be reasonable for
near equal mass systems at short periods where tidal synchronization
is expected to have occurred, but this is not a reasonable assumption
for \theobject, where only one out-of-eclipse modulation is seen.  The
observed rotational evolution of M-dwarfs
(e.g. \citealt{2011ApJ...727...56I}) indicates it is extremely
unlikely the two binary components could have the same rotation period
(and phase!) by chance in the absence of tidal effects, which are not
expected.

Unfortunately, at the time of writing, only a single primary eclipse
is available, so we are unable to take full advantage of the
non-synchronized spin and orbit at present.  Also, while multiple
secondary eclipses were observed, and residuals from our best-fitting
model (assuming no changes in eclipse shape or depth) are seen, it is
not clear if many of these are simply the result of systematic errors
in the photometry, given that equally large deviations are seen out of
eclipse.  Constraining the spot properties by this method will be an
important area for future work, and observing in multiple bandpasses
would be advantageous as it may provide information on the spot
temperatures.

\subsubsection{Spot model}

Conventionally, a Roche model such as the one implemented in the
popular Wilson-Devinney program \citep{1971ApJ...166..605W} would be
used to model a system with spots as these usually include a circular
spot model and perform the necessary surface integrals over the stars;
however the treatment of proximity effects is completely unnecessary
in the present system, and these models are extremely computationally
intensive, especially for systems with eccentric orbits, which would
make a detailed Monte Carlo based error analysis prohibitive.

It is also not clear that the circular spot model with a small number
of spots is realistic for late-type dwarfs.  Observed light curves
very rarely show the characteristic ``eclipse-like'' features with
flat baseline at maximum light, as would be predicted for a small
number of near-equatorial spots.  This indicates either that these
objects have very large, polar spots
(e.g. \citealt{1986A&A...165..135R}), such that some of the spot is
always in view to the observer to produce the continuous photometric
modulations seen in light curves, or that the surfaces of these
objects have many small spots
(e.g.
\citealt{2001MNRAS.326..950B,2004MNRAS.352..589B})
with the photometric modulations arising from a longitudinal
inhomogeneity of the spot distribution.  Modeling an ensemble of
spots, as in the latter case, in detail would be prohibitive, as even
single spot models are usually degenerate given limited light curve
data.  In our case, the degeneracies in fitting spot models would be
further exacerbated by the availability of only single-band light
curve information, meaning spot temperature and size would be
essentially degenerate.

Given these difficulties, we take an alternative approach for the
solution presented in this paper.  Rather than trying to model the
unknown spot distribution in detail, we simply attempt to incorporate
the effect of spots into the uncertainties on the final parameters.
We consider two models: (a) spots on the non-eclipsed part of the
photosphere on both stars; and (b), a case intended to approximate the
effect of spots covering the whole photosphere, again for both stars.
In both cases, the models are required to reproduce the observed out
of eclipse modulation.  It is necessary to run two models in each
spotted case because it is not known which star the out of eclipse
modulation originates from.

We assume a simple sinusoidal form for the modulations, which appears
to be an adequate description of the available light curve data.  The
functional form adopted was:
\begin{equation}
{\Delta L_i\over{L_i}} = a_i \sin\left({2 \pi F_i t\over{P}}\right) +
b_i \cos\left({2 \pi F_i t\over{P}}\right) - \sqrt{a_i^2+b_i^2}
\label{ooe_eq}
\end{equation}
where $t$ is time, $L_i$ is the light from star $i$, $a_i$ and $b_i$
are constants expressing the amplitude and phase of the modulation,
and $F_i$ is the ``rotation parameter'', the ratio of the rotation
frequency to the orbital frequency.  This expression is constructed to
yield $\Delta L_i = 0$ at maximum light, corresponding to the
conventional assumption that no spots are visible when the star is
brightest, which gives the minimum spot coverage necessary to
reproduce the observed modulations.

{\sc jktebop} computes the final system light (and thus the change in
magnitude) by summing the out of eclipse light and then subtracting
the eclipsed light.  By modulating only the out of eclipse light in
this summation using Eq. (\ref{ooe_eq}), hypothesis (a) can be
implemented, and hypothesis (b) can be implemented by also modulating
the eclipsed light (changing the surface brightness under the eclipse
chord) when the eclipsed star is star $i$.

\subsubsection{Results}
\label{spot-results-sect}

\begin{figure}
\centering
\includegraphics[angle=270,width=3.1in]{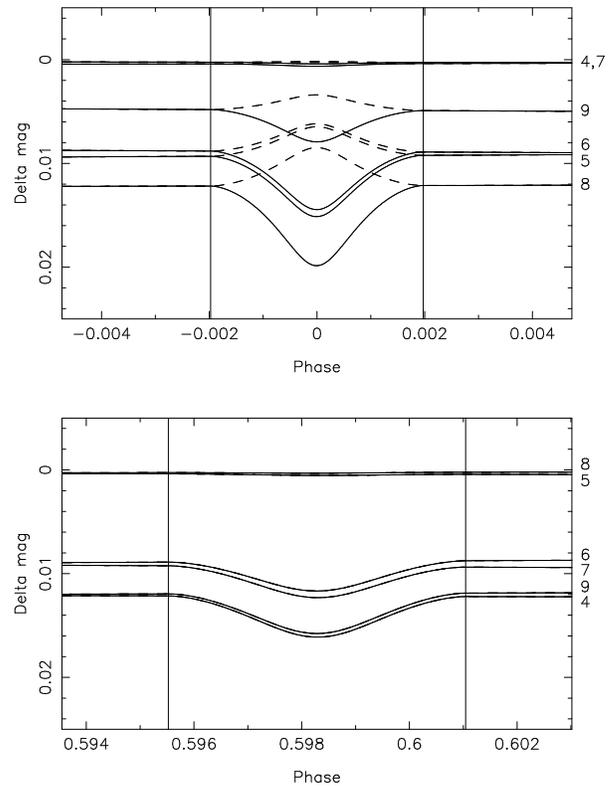}
\caption{Predicted deviation in the primary (top) and secondary
  (bottom) eclipses resulting from spots on the primary star.  The
  solid curves correspond to hypothesis (a) where the spots are not
  eclipsed, and the dashed curves to (b) where they are eclipsed.  The
  two sets of curves are identical in the lower panel.  The cycle
  numbers (integer part of the normalized orbital phase) are shown on
  the right of each panel.}
\label{resid-spot1}
\end{figure}

\begin{figure}
\centering
\includegraphics[angle=270,width=3.1in]{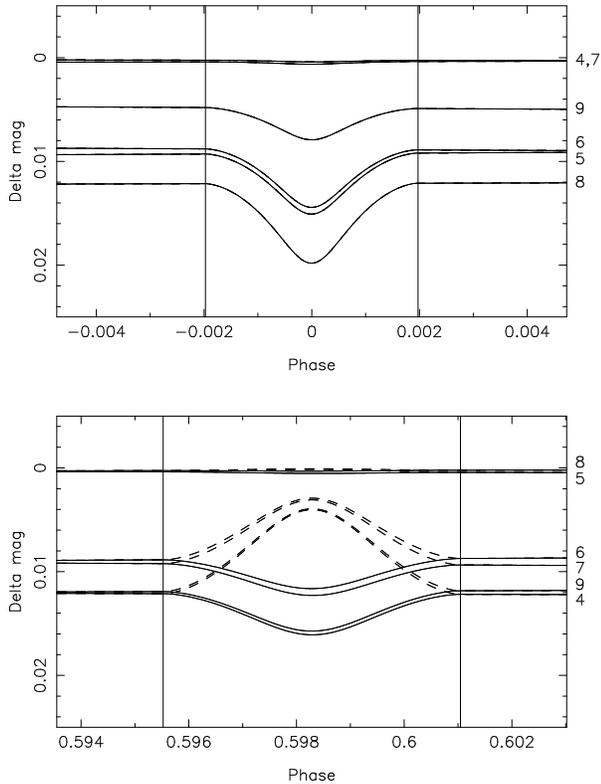}
\caption{Predicted deviation in the primary (top) and secondary
  (bottom) eclipses resulting from spots on the secondary star.
  Curves as Figure \ref{resid-spot1}.  The two sets of curves are
  identical in the upper panel.}
\label{resid-spot2}
\end{figure}

The effect of applying our spot model on the eclipses is shown in
Figures \ref{resid-spot1} and \ref{resid-spot2}, which were calculated
using the parameters of \theobject, comparing a model with spots to a
model with identical physical parameters, but without spots.  This
figure demonstrates that observing multiple eclipses at high
precision would allow the star hosting the spots to be identified.

\begin{figure*}
\centering
\includegraphics[angle=270,width=7in]{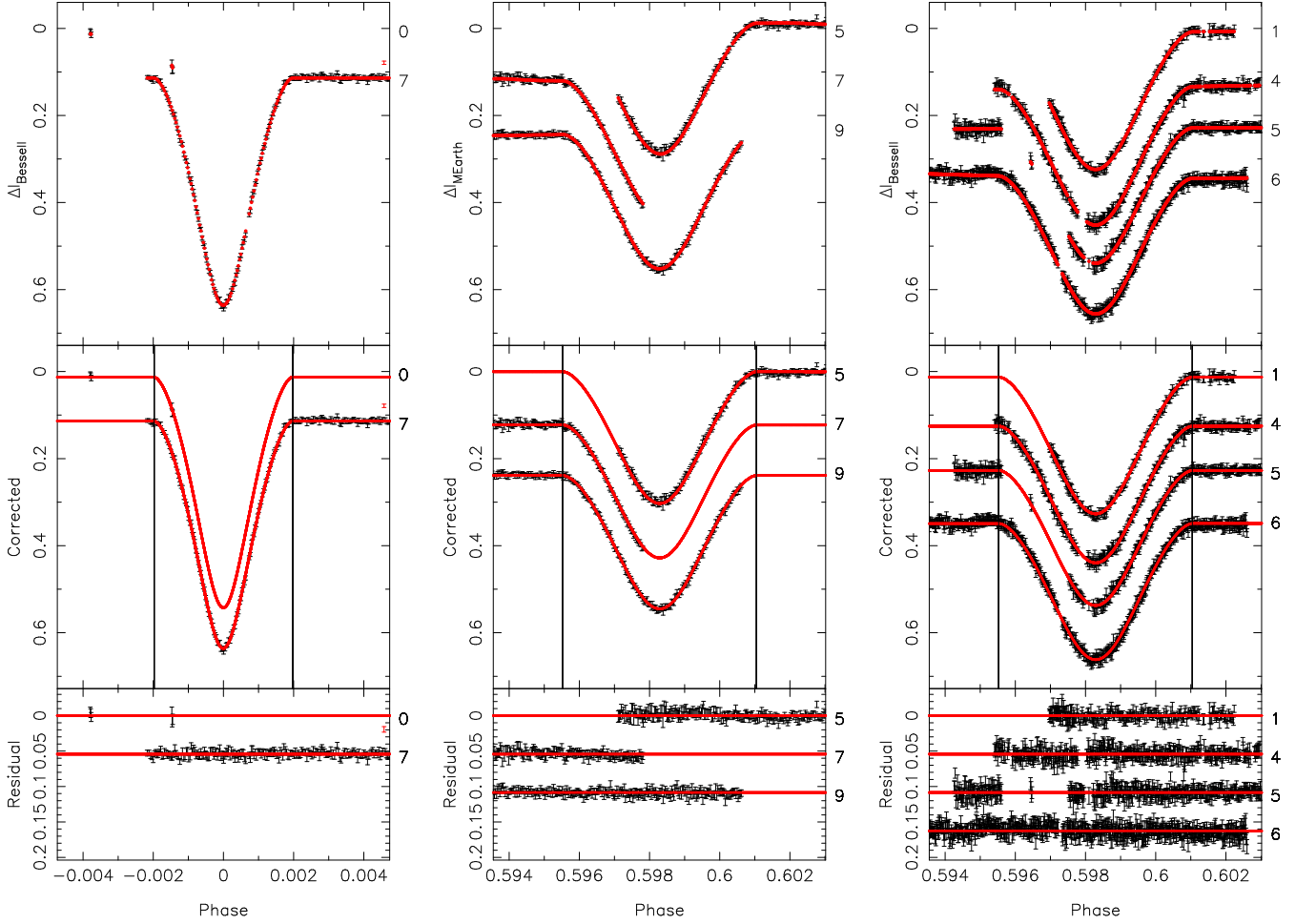}
\caption{Eclipse light curves used in the model, with the best fit
  overlaid (the model is the first of the assumptions in Table
  \ref{fit-spot-params}, but the other model light curves are almost
  identical in appearance).  In the top panels, the model is shown
  with red dots at the epochs of the observations, and in the other
  panels it is plotted as a continuous curve.  The figure has been
  divided horizontally into one column for each combination of eclipse
  and passband, with the primary eclipses (all in $\ibess$) plotted on
  the left, the secondary eclipses in $\imear$ in the center, and in
  $\ibess$ on the right.  Each column contains three panels showing
  the raw differential photometry (top), the same after correction to
  flatten the out of eclipse baseline (center; these corrections are
  the airmass term for the Clay and MEarth secondary eclipses, and the
  ``common-mode'' term for the MEarth secondary eclipses) and the
  residual in the bottom panel.  We have offset the different eclipses
  vertically for clarity, and the cycle number (integer part of the
  normalized orbital phase) is shown on the right.  The vertical bars
  indicate the approximate locations of the first and last contact
  points.}
\label{lc-fit}
\end{figure*}

\begin{figure*}
\centering
\includegraphics[angle=270,width=7in]{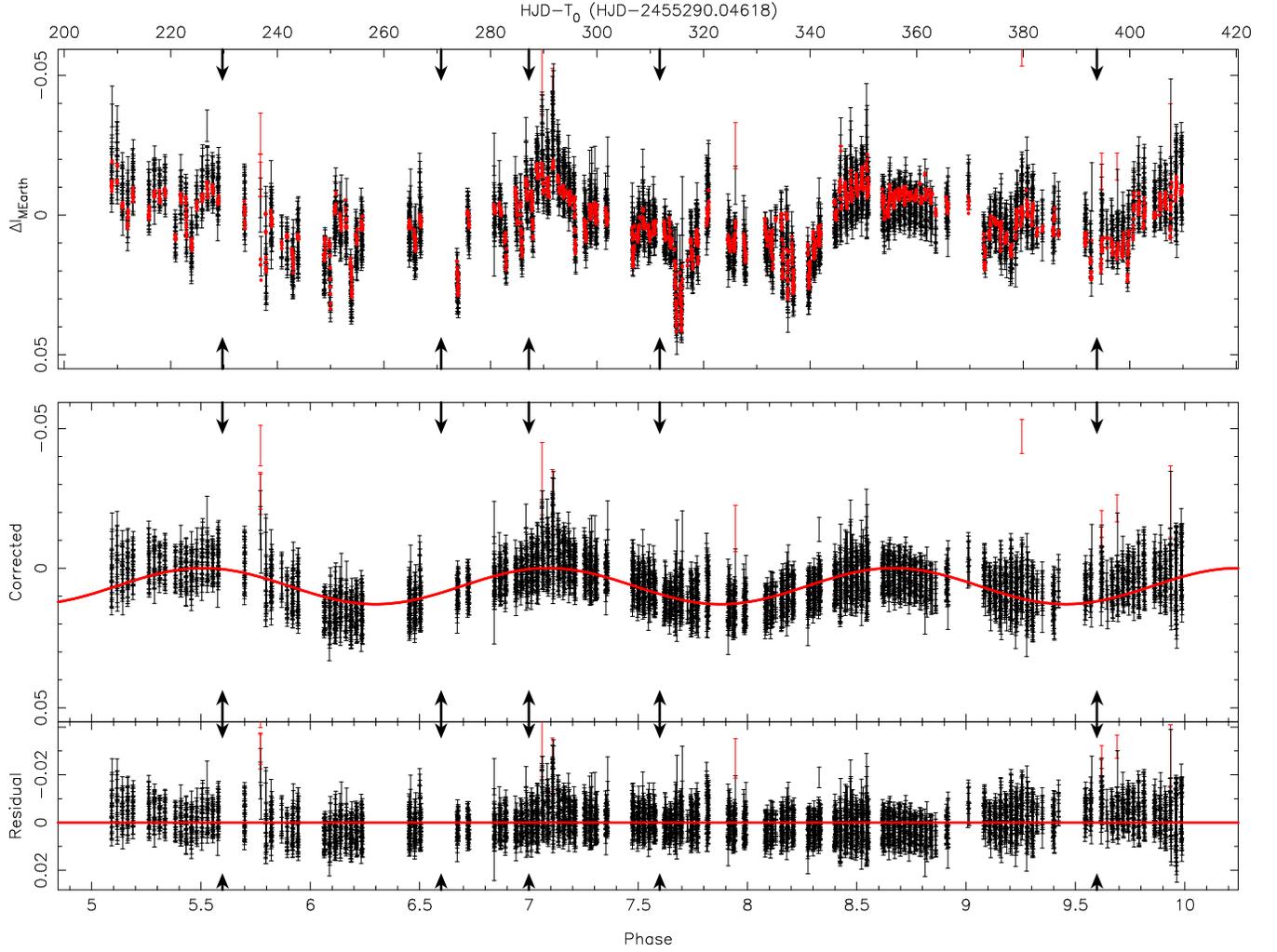}
\caption{MEarth 2010-2011 season out of eclipse light curve.  Vertical
  panels as Fig. \ref{lc-fit}, with the best-fitting sinusoidal model
  overplotted.  Arrows indicate the phases corresponding to the
  observed eclipses.}
\label{ooe-fit}
\end{figure*}

\begin{figure*}
\centering
\includegraphics[angle=270,width=7in]{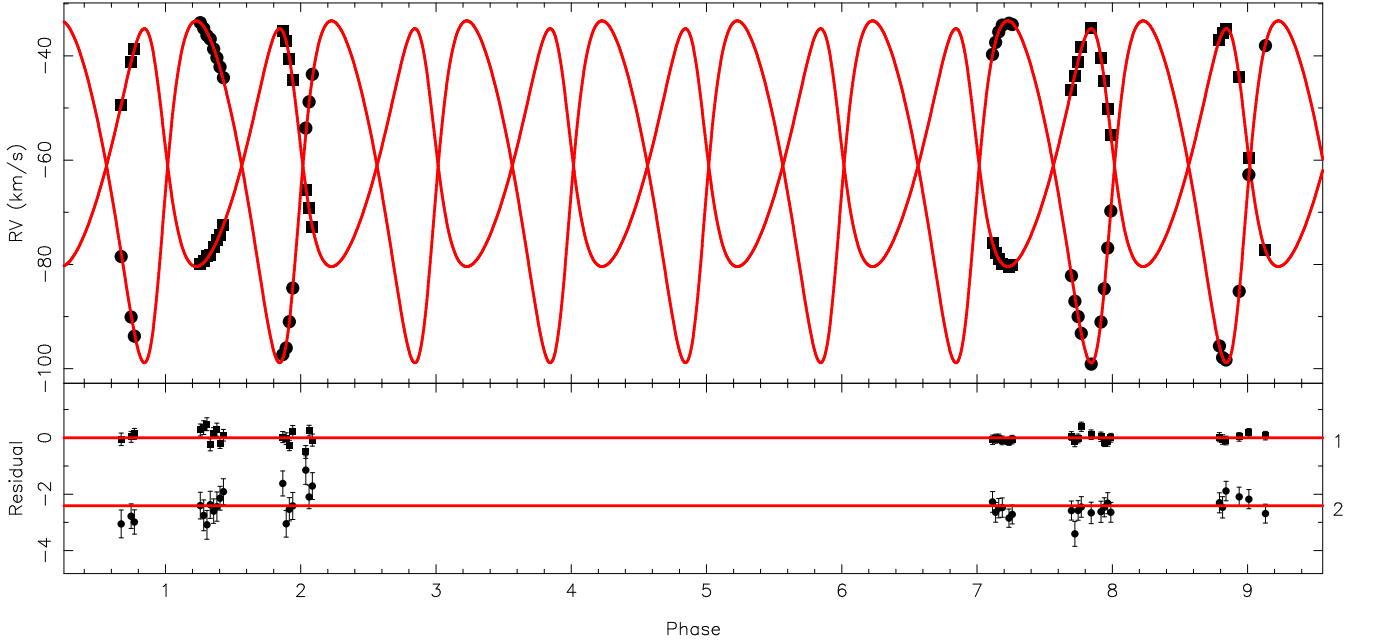}
\caption{Radial velocity curves used in the model, with the best fit
  overlaid (top) and residual (bottom).  The primary velocities are
  shown with square symbols, and the secondary velocities with round
  symbols.  The residuals for the two stars have been offset for
  clarity.}
\label{rv-fit}
\end{figure*}

We fit all four spotted models to the full data-set for \theobject\
using the procedures already described.  The results are given in
Table \ref{fit-spot-params}, and Figures \ref{lc-fit}, \ref{ooe-fit}
and \ref{rv-fit} show the data with a representative model,
corresponding to hypothesis (a), the ``non-eclipsed spots'' model, on
the primary, overplotted.

\begin{deluxetable*}{lllll}
\tabletypesize{\normalsize}
\tablecaption{\label{fit-spot-params} Derived parameters and
  uncertainties for the four spot configurations with no additional spots.}
\tablecolumns{5}

\tablehead{
 & \multicolumn{2}{c}{Non-eclipsed spots} & \multicolumn{2}{c}{Eclipsed spots} \\
\colhead{Parameter\tablenotemark{a}} & \colhead{Primary} & \colhead{Secondary} & \colhead{Primary} & \colhead{Secondary}
}

\startdata
$J_{\rm MEarth}$          &                $0.7999 \pm 0.0064$ &                $0.8004 \pm 0.0066$ &                $0.7996 \pm 0.0064$ &                $0.8423 \pm 0.0047$ \\
$J_{\rm Bessell}$         &                $0.8255 \pm 0.0066$ &                $0.8259 \pm 0.0069$ &                $0.8252 \pm 0.0066$ &                $0.8637 \pm 0.0043$ \\
$(R_1+R_2)/a$             & $0.015548^{+0.000039}_{-0.000042}$ & $0.015551^{+0.000042}_{-0.000044}$ & $0.015550^{+0.000041}_{-0.000042}$ & $0.015661^{+0.000025}_{-0.000026}$ \\
$R_2/R_1$                 &                  $0.787 \pm 0.017$ &                  $0.788 \pm 0.018$ &                  $0.788 \pm 0.017$ &                  $0.888 \pm 0.020$ \\
$\cos i$                  & $0.004645^{+0.000103}_{-0.000112}$ & $0.004652^{+0.000109}_{-0.000115}$ & $0.004646^{+0.000105}_{-0.000111}$ & $0.005005^{+0.000051}_{-0.000058}$ \\
$P$ (days)                &           $41.032363 \pm 0.000024$ &           $41.032364 \pm 0.000024$ &           $41.032365 \pm 0.000024$ &           $41.032336 \pm 0.000025$ \\
$T_0$ (HJD\_UTC)          &        $2455290.04618 \pm 0.00018$ &        $2455290.04618 \pm 0.00018$ &        $2455290.04617 \pm 0.00018$ &        $2455290.04637 \pm 0.00018$ \\
 \\
$F_1$                     &                $0.6332 \pm 0.0019$ &                             \ldots &                $0.6331 \pm 0.0019$ &                             \ldots \\
$a_1$ (mag)               &              $0.00602 \pm 0.00067$ &                             \ldots &              $0.00599 \pm 0.00067$ &                             \ldots \\
$b_1$ (mag)               &              $0.00629 \pm 0.00062$ &                             \ldots &              $0.00632 \pm 0.00062$ &                             \ldots \\
$F_2$                     &                             \ldots &                $0.6332 \pm 0.0019$ &                             \ldots &                $0.6327 \pm 0.0020$ \\
$a_2$ (mag)               &                             \ldots &                $0.0125 \pm 0.0014$ &                             \ldots &                $0.0098 \pm 0.0012$ \\
$b_2$ (mag)               &                             \ldots &                $0.0131 \pm 0.0013$ &                             \ldots &                $0.0109 \pm 0.0011$ \\
 \\
$s_{\rm OOE}$\tablenotemark{b}             &                  $1.469 \pm 0.014$ &                  $1.467 \pm 0.014$ &                  $1.467 \pm 0.014$ &                  $1.467 \pm 0.014$ \\
$s_{\rm MEarth-20101119}$ &                  $1.191 \pm 0.058$ &                  $1.192 \pm 0.057$ &                  $1.191 \pm 0.058$ &                  $1.498 \pm 0.086$ \\
$s_{\rm MEarth-20110209}$ &                  $1.548 \pm 0.087$ &                  $1.548 \pm 0.087$ &                  $1.549 \pm 0.087$ &                  $1.577 \pm 0.089$ \\
$s_{\rm MEarth-20110502}$ &                  $1.194 \pm 0.055$ &                  $1.194 \pm 0.056$ &                  $1.193 \pm 0.055$ &                  $1.260 \pm 0.064$ \\
$s_{\rm MEarth-Primary}$  &                    $1.40 \pm 0.36$ &                    $1.39 \pm 0.35$ &                    $1.42 \pm 0.36$ &                    $1.39 \pm 0.35$ \\
$s_{\rm Hankasalmi}$      &                  $1.135 \pm 0.049$ &                  $1.136 \pm 0.049$ &                  $1.137 \pm 0.049$ &                  $1.143 \pm 0.050$ \\
$s_{\rm Clay-20100608}$   &                  $1.022 \pm 0.046$ &                  $1.023 \pm 0.046$ &                  $1.022 \pm 0.047$ &                  $1.034 \pm 0.046$ \\
$s_{\rm Clay-20101009}$   &                  $1.457 \pm 0.048$ &                  $1.455 \pm 0.048$ &                  $1.456 \pm 0.050$ &                  $1.498 \pm 0.050$ \\
$s_{\rm Clay-20101119}$   &                  $1.177 \pm 0.038$ &                  $1.179 \pm 0.038$ &                  $1.178 \pm 0.038$ &                  $1.210 \pm 0.040$ \\
$s_{\rm Clay-20101230}$   &                  $1.367 \pm 0.034$ &                  $1.367 \pm 0.034$ &                  $1.367 \pm 0.034$ &                  $1.357 \pm 0.033$ \\
 \\
$k_{\rm MEarth}$          &             $-0.00659 \pm 0.00037$ &             $-0.00658 \pm 0.00037$ &             $-0.00659 \pm 0.00038$ &             $-0.00686 \pm 0.00039$ \\
$C_{\rm MEarth}$          &                $0.9834 \pm 0.0080$ &                $0.9841 \pm 0.0080$ &                $0.9842 \pm 0.0080$ &                $0.9868 \pm 0.0080$ \\
$k_{\rm Clay-20100608}$   &               $-0.0076 \pm 0.0031$ &               $-0.0074 \pm 0.0031$ &               $-0.0075 \pm 0.0031$ &               $-0.0204 \pm 0.0032$ \\
$k_{\rm Clay-20101009}$   &                $0.0136 \pm 0.0017$ &                $0.0135 \pm 0.0017$ &                $0.0135 \pm 0.0017$ &                $0.0184 \pm 0.0017$ \\
$k_{\rm Clay-20101119}$   &                $0.0042 \pm 0.0014$ &                $0.0041 \pm 0.0014$ &                $0.0041 \pm 0.0014$ &                $0.0031 \pm 0.0014$ \\
$k_{\rm Clay-20101230}$   &               $-0.0228 \pm 0.0014$ &               $-0.0228 \pm 0.0014$ &               $-0.0229 \pm 0.0014$ &               $-0.0244 \pm 0.0014$ \\
 \\
$e \cos \omega$           &            $0.152408 \pm 0.000033$ &            $0.152406 \pm 0.000034$ &            $0.152409 \pm 0.000034$ &            $0.152327 \pm 0.000035$ \\
$e \sin \omega$           &                $0.1823 \pm 0.0011$ &                $0.1824 \pm 0.0012$ &                $0.1823 \pm 0.0012$ &                $0.1852 \pm 0.0012$ \\
$q$                       &                $0.6958 \pm 0.0020$ &                $0.6958 \pm 0.0020$ &                $0.6958 \pm 0.0020$ &                $0.6958 \pm 0.0022$ \\
$(K_1+K_2)$ ($\kms$)      &                 $55.602 \pm 0.087$ &                 $55.602 \pm 0.087$ &                 $55.603 \pm 0.087$ &                 $55.601 \pm 0.091$ \\
$\gamma$ ($\kms$)\tablenotemark{c}         &                $-61.071 \pm 0.027 \pm 0.5$ &                $-61.071 \pm 0.027 \pm 0.5$ &                $-61.070 \pm 0.027 \pm 0.5$ &                $-61.068 \pm 0.028 \pm 0.5$ \\
$s_1$ ($\kms$)            &                  $0.125 \pm 0.015$ &                  $0.125 \pm 0.015$ &                  $0.125 \pm 0.015$ &                  $0.134 \pm 0.017$ \\
$s_2$ ($\kms$)            &                  $0.280 \pm 0.033$ &                  $0.279 \pm 0.032$ &                  $0.279 \pm 0.032$ &                  $0.290 \pm 0.034$ \\
 \\
$i$ ($^\circ$)            &      $89.7339^{+0.0064}_{-0.0059}$ &      $89.7335^{+0.0066}_{-0.0062}$ &      $89.7338^{+0.0063}_{-0.0061}$ &      $89.7132^{+0.0033}_{-0.0029}$ \\
$e$                       &              $0.23763 \pm 0.00086$ &              $0.23770 \pm 0.00088$ &              $0.23762 \pm 0.00087$ &              $0.23977 \pm 0.00090$ \\
$\omega$ ($^\circ$)       &                   $50.10 \pm 0.18$ &                   $50.12 \pm 0.19$ &                   $50.10 \pm 0.19$ &                   $50.56 \pm 0.19$ \\
$a$ ($\rsol$)             &                 $43.825 \pm 0.069$ &                 $43.824 \pm 0.070$ &                 $43.825 \pm 0.070$ &                 $43.800 \pm 0.073$ \\
$L_2/L_1$                 &                  $0.485 \pm 0.025$ &                  $0.474 \pm 0.025$ &                  $0.486 \pm 0.025$ &                  $0.636 \pm 0.031$ \\
$T_{\rm sec}$ (HJD\_UTC)  &        $2455314.59517 \pm 0.00015$ &        $2455314.59516 \pm 0.00015$ &        $2455314.59516 \pm 0.00015$ &        $2455314.59536 \pm 0.00015$ \\
 \\
$M_1$ ($\msol$)           &                $0.3951 \pm 0.0022$ &                $0.3950 \pm 0.0022$ &                $0.3951 \pm 0.0022$ &                $0.3944 \pm 0.0023$ \\
$M_2$ ($\msol$)           &                $0.2749 \pm 0.0011$ &                $0.2749 \pm 0.0011$ &                $0.2749 \pm 0.0011$ &                $0.2744 \pm 0.0012$ \\
$(R_1+R_2)$ ($\rsol$)     &       $0.6814^{+0.0020}_{-0.0020}$ &       $0.6815^{+0.0021}_{-0.0021}$ &       $0.6815^{+0.0020}_{-0.0021}$ &       $0.6860^{+0.0016}_{-0.0016}$ \\
$R_1$ ($\rsol$)           &       $0.3815^{+0.0028}_{-0.0030}$ &       $0.3814^{+0.0028}_{-0.0032}$ &       $0.3814^{+0.0028}_{-0.0031}$ &       $0.3635^{+0.0035}_{-0.0037}$ \\
$R_2$ ($\rsol$)           &       $0.2999^{+0.0044}_{-0.0043}$ &       $0.3001^{+0.0047}_{-0.0044}$ &       $0.3001^{+0.0045}_{-0.0043}$ &       $0.3225^{+0.0043}_{-0.0042}$ \\
\enddata

\tablenotetext{a}{Parameter names as defined in Table \ref{model_pars} and in the text.}
\tablenotetext{b}{The acronym OOE is used as shorthand for ``out of eclipse''.}
\tablenotetext{c}{Errors quoted for the $\gamma$ velocity are the
  random error (scatter), followed by our estimated systematic error
  in the velocity zero-point, from the assumed barycentric velocity of
  Barnard's star.} 

\end{deluxetable*}

\begin{figure}
\centering
\includegraphics[angle=270,width=2.2in]{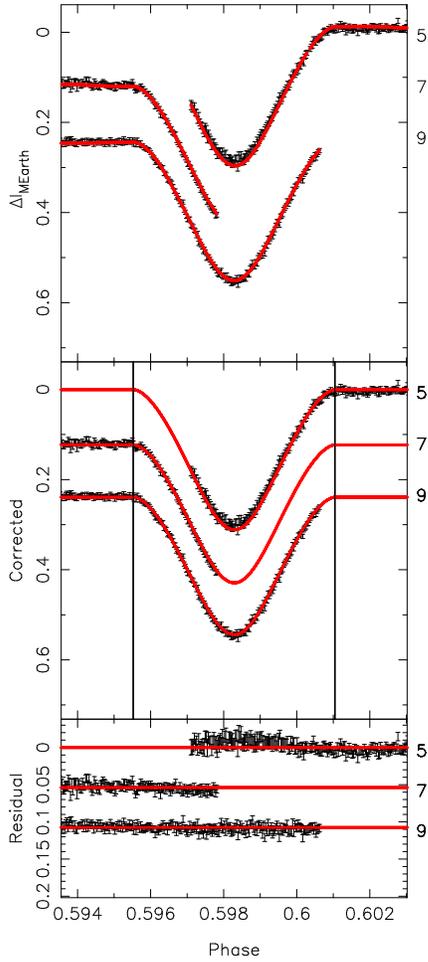}
\caption{As Figure \ref{lc-fit}, except for the ``eclipsed spots''
  model on the secondary, showing the deviations in the MEarth 2010
  November 19 event.}
\label{lc-fit-ecl2}
\end{figure}

The 2010 November 19 partial secondary eclipse from MEarth (numbered
$5$ in the figure) provides the most discriminating power between the
various models, particularly taken in conjunction with the 2011 May 2
event (numbered $9$), which dominates the fit.  As shown in Figure
\ref{ooe-fit}, these were taken at opposite extremes of the out of
eclipse variation, with the 2010 November 19 event at maximum light
and the 2011 May 2 event close to minimum light.  Examining the sizes
of the $s$ parameter in Table \ref{fit-spot-params}, the ``eclipsed
spots'' model on the secondary, which is the only model presenting a
significant detectable deviation in the secondary eclipses (as shown
in Figure \ref{resid-spot2}) is disfavored by the data (see Figure
\ref{lc-fit-ecl2}), having an $s_{\rm MEarth-20101119}$ value larger
by almost $3 \sigma$ than the ``non-eclipsed spots'' model on the
secondary.  Indeed, the observations indicate that this eclipse needs
to be made as shallow as possible in the model in order to produce a
good fit, possibly even slightly shallower than the ``non-eclipsed
spots'' model is able to produce (the discrepancy may arise from the
imperfect modeling of the out-of-eclipse variation by a sinusoid).
The depth of this eclipse therefore argues that if the spots causing
the out of eclipse modulation are on the secondary, they should be at
latitudes not crossed by the eclipse chord.

It is interesting to note that, because the radius ratio is quite
close to unity, the range of allowed combinations of spot latitude and
inclination angle for the secondary that would explain the observed
out of eclipse modulations is therefore quite narrow, compared to the
much wider range of possibilities on the primary that are still
allowed.  This mildly argues in favor of spots on the primary as being
the more likely explanation, purely from geometry.

Nevertheless, with the presently-available data, the other models do
not show any clear disagreement with the observations, so we consider
(from a purely observational point of view) that all three are equally
likely at present.

\subsubsection{Effect of additional spots}
\label{add_spot_sect}

The model we have considered so far uses the minimum spot coverage
necessary to reproduce the observed out-of-eclipse modulations.  As
noted, e.g. by \citet{2011MNRAS.412.1599B} and other authors, studies
of cool stars sensitive to the total spot coverage typically find much
higher filling factors than photometric spot models.  Therefore it is
likely these stars have a significant longitudinally homogeneous spot
component.

In this section, we consider the effect of adding additional spots in a
longitudinally homogeneous fashion where they would not be detectable
through modulations in the photometry.  For simplicity, we have placed
these spots on the same star and at the same latitudes as the
longitudinally inhomogeneous component, noting that we are
predominantly interested in the effect on the component radii, where
it does not matter which of the possible locations is used for the
longitudinally inhomogeneous spot component.

As discussed earlier, the spectroscopic light ratio is sensitive to
differences in the overall spot coverage between the two stars in some
cases.  We have explored this limit, and the effect of adding varying
quantities of spots on the component radii, by subtracting an extra
term $c_i$ in Eq. (\ref{ooe_eq}) representing the fraction of stellar
light removed by the longitudinally homogeneous spot component.

For single late-type active dwarfs, it is typical to find spot
filling factors of $20-50\%$ from observations sensitive to the entire
surface spot coverage (e.g. \citealt{2004AJ....128.1802O}), although
we note the available information for M-dwarfs is very limited.
Photometric observations of M-dwarfs indicate spot temperature
contrast of up to $10\%$ (e.g. \citealt{2006A&A...448.1111R}, who
find $4-8\%$).  This yields a decrement of approximately $10-20\%$ in
the $I$-band stellar light, assuming parameters appropriate to the
present system, with the lower end of the range presumably more
typical for less active stars.  We therefore consider values of $c_i =
0.1$ and $0.2$.  The latter corresponds approximately to the spot
levels discussed by \citet{2010ApJ...718..502M} for the existing
short-period eclipsing binaries.  The results are given in Tables
\ref{fit-extspot1-params} and \ref{fit-extspot2-params}.

\begin{deluxetable*}{lllllll}
\tabletypesize{\normalsize}
\tablecaption{\label{fit-extspot1-params} Derived parameters and
  uncertainties showing the effect of additional spots on the
  primary.}
\tablecolumns{7}

\tablehead{
 & \multicolumn{3}{c}{Non-eclipsed spots} & \multicolumn{3}{c}{Eclipsed spots} \\
\colhead{Parameter} & \colhead{$c_1 = 0$} & \colhead{$c_1 = 0.1$} &
\colhead{$c_1 = 0.2$} & \colhead{$c_1 = 0$} & \colhead{$c_1 = 0.1$} &
\colhead{$c_1 = 0.2$}
}

\startdata
$J_{\rm MEarth}$          &                $0.7999 \pm 0.0064$ &                $0.8302 \pm 0.0066$ &                $0.8724 \pm 0.0074$ &                $0.7996 \pm 0.0064$ &                $0.7196 \pm 0.0060$ &                $0.6396 \pm 0.0051$ \\
$J_{\rm Bessell}$         &                $0.8255 \pm 0.0066$ &                $0.8583 \pm 0.0068$ &                $0.9042 \pm 0.0079$ &                $0.8252 \pm 0.0066$ &                $0.7427 \pm 0.0062$ &                $0.6602 \pm 0.0054$ \\
$(R_1+R_2)/a$             & $0.015548^{+0.000039}_{-0.000042}$ & $0.015632^{+0.000039}_{-0.000039}$ & $0.015775^{+0.000037}_{-0.000039}$ & $0.015550^{+0.000041}_{-0.000042}$ & $0.015551^{+0.000042}_{-0.000044}$ & $0.015550^{+0.000041}_{-0.000042}$ \\
$R_2/R_1$                 &                  $0.787 \pm 0.017$ &                  $0.766 \pm 0.016$ &                  $0.756 \pm 0.017$ &                  $0.788 \pm 0.017$ &                  $0.788 \pm 0.018$ &                  $0.787 \pm 0.017$ \\
$\cos i$                  & $0.004645^{+0.000103}_{-0.000112}$ & $0.005219^{+0.000083}_{-0.000086}$ & $0.005861^{+0.000065}_{-0.000072}$ & $0.004646^{+0.000105}_{-0.000111}$ & $0.004646^{+0.000109}_{-0.000116}$ & $0.004644^{+0.000107}_{-0.000112}$ \\
\\
$i$ ($^\circ$)            &      $89.7339^{+0.0064}_{-0.0059}$ &      $89.7010^{+0.0049}_{-0.0047}$ &      $89.6642^{+0.0041}_{-0.0037}$ &      $89.7338^{+0.0063}_{-0.0061}$ &      $89.7338^{+0.0067}_{-0.0063}$ &      $89.7339^{+0.0064}_{-0.0061}$ \\
$L_2/L_1$                 &                  $0.485 \pm 0.025$ &                  $0.531 \pm 0.025$ &                  $0.611 \pm 0.032$ &                  $0.486 \pm 0.025$ &                  $0.486 \pm 0.026$ &                  $0.486 \pm 0.025$ \\
\\
$(R_1+R_2)$ ($\rsol$)     &       $0.6814^{+0.0020}_{-0.0020}$ &       $0.6847^{+0.0020}_{-0.0020}$ &       $0.6904^{+0.0020}_{-0.0021}$ &       $0.6815^{+0.0020}_{-0.0021}$ &       $0.6815^{+0.0021}_{-0.0021}$ &       $0.6815^{+0.0020}_{-0.0021}$ \\
$R_1$ ($\rsol$)           &       $0.3815^{+0.0028}_{-0.0030}$ &       $0.3878^{+0.0026}_{-0.0029}$ &       $0.3934^{+0.0030}_{-0.0033}$ &       $0.3814^{+0.0028}_{-0.0031}$ &       $0.3814^{+0.0029}_{-0.0032}$ &       $0.3814^{+0.0028}_{-0.0031}$ \\
$R_2$ ($\rsol$)           &       $0.2999^{+0.0044}_{-0.0043}$ &       $0.2969^{+0.0042}_{-0.0039}$ &       $0.2970^{+0.0045}_{-0.0043}$ &       $0.3001^{+0.0045}_{-0.0043}$ &       $0.3001^{+0.0047}_{-0.0044}$ &       $0.3000^{+0.0045}_{-0.0043}$ \\
\enddata

\end{deluxetable*}

\begin{deluxetable*}{llll}
\tabletypesize{\normalsize}
\tablecaption{\label{fit-extspot2-params} Derived parameters and
  uncertainties showing the effect of additional spots on the secondary.}
\tablecolumns{4}

\tablehead{
 & \multicolumn{3}{c}{Non-eclipsed spots} \\
\colhead{Parameter} & \colhead{$c_2 = 0$} & \colhead{$c_2 = 0.1$} &
 \colhead{$c_2 = 0.2$}
}

\startdata
$J_{\rm MEarth}$          &                $0.8004 \pm 0.0066$ &                $0.8157 \pm 0.0075$ &                $0.8271 \pm 0.0081$ \\
$J_{\rm Bessell}$         &                $0.8259 \pm 0.0069$ &                $0.8409 \pm 0.0077$ &                $0.8516 \pm 0.0084$ \\
$(R_1+R_2)/a$             & $0.015551^{+0.000042}_{-0.000044}$ & $0.015586^{+0.000043}_{-0.000044}$ & $0.015601^{+0.000045}_{-0.000045}$ \\
$R_2/R_1$                 &                  $0.788 \pm 0.018$ &                  $0.777 \pm 0.016$ &                  $0.757 \pm 0.014$ \\
$\cos i$                  &    $0.00465^{+0.00011}_{-0.00012}$ &    $0.00493^{+0.00011}_{-0.00012}$ &    $0.00513^{+0.00012}_{-0.00012}$ \\
\\
$i$ ($^\circ$)            &      $89.7335^{+0.0066}_{-0.0062}$ &      $89.7177^{+0.0066}_{-0.0064}$ &      $89.7063^{+0.0067}_{-0.0066}$ \\
$L_2/L_1$                 &                  $0.474 \pm 0.025$ &                  $0.423 \pm 0.022$ &                  $0.360 \pm 0.017$ \\
\\
$(R_1+R_2)$ ($\rsol$)     &       $0.6815^{+0.0021}_{-0.0021}$ &       $0.6829^{+0.0021}_{-0.0021}$ &       $0.6834^{+0.0022}_{-0.0022}$ \\
$R_1$ ($\rsol$)           &       $0.3814^{+0.0028}_{-0.0032}$ &       $0.3844^{+0.0026}_{-0.0029}$ &       $0.3892^{+0.0021}_{-0.0024}$ \\
$R_2$ ($\rsol$)           &       $0.3001^{+0.0047}_{-0.0044}$ &       $0.2985^{+0.0044}_{-0.0042}$ &       $0.2942^{+0.0040}_{-0.0037}$ \\
\enddata

\end{deluxetable*}

Firstly, it is clear from Table \ref{fit-extspot1-params} that adding
additional ``eclipsed spots'' has no effect on the radii or light
ratio and merely alters the surface brightness ratio (this means it
also changes the effective temperature ratio).  This is also true when
placing these spots on the secondary, but we do not show results for
this model since it is unable to reproduce the observed eclipse light
curves, as discussed in \S \ref{spot-results-sect}.

Adding ``non-eclipsed spots'' affects the light ratio, and both
component radii, slightly increasing the sum of the radii, increasing
$R_1$, and decreasing $R_2$.  The light ratio behaves as expected,
becoming closer to unity as the primary is darkened by the addition of
spots.  Both $c_2 = 0.2$ models with ``non-eclipsed spots'' (and to
some extent, the $c_2 = 0.1$ model) produce a light ratio that is
marginally discrepant with the spectroscopic value (see \S
\ref{spec_sect} or Table \ref{model_pars}), but the other models agree
reasonably well within the errors.

Given the lack of clearly-detected perturbations in the best quality
eclipse curves, it is reasonable to favor the ``non-eclipsed spots''
models over the ``eclipsed spots'' models.  Assuming the $c_i = 0.1$
model is typical, and the $c_i = 0$ and $0.2$ models represent the
likely range of values, we estimate the final system parameters and
their uncertainties by combining the posterior probability
distributions from all six ``non-eclipsed spots'' models in this
section (all three values of $c_i$).  These parameters and estimated
uncertainties are reported in \S \ref{mod_sum_sect}.

\subsection{Bandpass mismatch}
\label{mod_band_sect}

A concern when combining light curves taken with different
instruments, is the effect of any difference in the photometric
bandpasses.  In the present case, where the primary and secondary
eclipses were (by necessity) measured using different instruments,
this predominantly affects the ratio of the eclipse depths, and thus
the parameter $J$, the ratio of central surface brightnesses.

In order to determine the approximate size of the bandpass mismatch,
we measured the difference in magnitude between \theobject\ and a
nearby, much bluer comparison star, at $11^h12^m12^s.15$
$+76^\circ27\arcmin33\farcs8$ (position from 2MASS; this is the bright
star seen near the upper left corner of Figure
\ref{third_light_images}).  The latter star has $(J-K)_{\rm 2MASS} =
0.27$ and is thus probably in the F spectral class.  It is
non-variable at the precision of the MEarth data (rms scatter
$0.003\ {\rm mag}$).

Our measured magnitude differences were $\Delta I({\rm Hankasalmi}) =
0.36$, $\Delta I({\rm Clay}) = 0.35$, and $\Delta I({\rm MEarth}) =
0.55$.  The excellent agreement between the two $\ibess$ filters
(especially in light of the observed out of eclipse variations of our
target) indicates they are most likely an extremely good match.

Assuming a simple linear scaling, we estimate the change in $J$
corresponding to the mismatch between the two $\ibess$ filters is
approximately $1/10$ of that between the $\ibess$ and $\imear$
filters, or $\delta J \approx 0.003$.  This is unimportant compared to
the other sources of uncertainty (see Table \ref{fit-spot-params}).

\subsection{Limb darkening}
\label{mod_ld_sect}

\subsubsection{Differences between $\ibess$ and $\icous$}

We first examine the effect of assuming the limb darkening law is the
same in the two $I$ filters.  To do this, we compare the fitting
results using our baseline model to using the coefficients for
the $z'$ SDSS passband from \citet{2004A&A...428.1001C}, where the
$\ibess$ band should lie inbetween these two extremes.

Using the observed colors of \theobject, we estimate that
approximately $1/3$ the difference between the $I_C$ and $z'$ results
is appropriate for the error introduced by our assumption of the
$\icous$ limb darkening law in fitting the $\ibess$ data.  Comparing
the results for the two bands in Table \ref{fit-zld-params}, we find
this source of error in the radii is negligible (although it does
affect $J$ and $i$ at close to $1 \sigma$).  While the radius
sum error is comparable to the individual estimated uncertainties in
the table, the uncertainty in this parameter for our adopted solution
combining six spot configurations is larger, and the limb darkening
contribution is then less important.

\begin{deluxetable}{lll}
\tabletypesize{\normalsize}
\tablecaption{\label{fit-zld-params} Comparison of derived parameters
  and uncertainties for $\icous$ and $z'$ limb darkening laws using
  the ``non-eclipsed spots'' model on the primary.}
\tablecolumns{3}

\tablehead{
\colhead{Parameter} & \colhead{$\icous$} & \colhead{$z'$}
}

\startdata
$J_{\rm MEarth}$          &                $0.7999 \pm 0.0064$ &                $0.7829 \pm 0.0085$ \\
$J_{\rm Bessell}$         &                $0.8255 \pm 0.0066$ &                $0.8073 \pm 0.0091$ \\
$(R_1+R_2)/a$             & $0.015548^{+0.000039}_{-0.000042}$ & $0.015387^{+0.000053}_{-0.000056}$ \\
$R_2/R_1$                 &                  $0.787 \pm 0.017$ &                  $0.773 \pm 0.022$ \\
$\cos i$                  &    $0.00464^{+0.00010}_{-0.00011}$ &    $0.00440^{+0.00016}_{-0.00017}$ \\
\\
$i$ ($^\circ$)            &      $89.7339^{+0.0064}_{-0.0059}$ &      $89.7481^{+0.0096}_{-0.0089}$ \\
$L_2/L_1$                 &                  $0.485 \pm 0.025$ &                  $0.462 \pm 0.031$ \\
 \\
$(R_1+R_2)$ ($\rsol$)     &       $0.6814^{+0.0020}_{-0.0020}$ &       $0.6744^{+0.0025}_{-0.0027}$ \\
$R_1$ ($\rsol$)           &       $0.3815^{+0.0028}_{-0.0030}$ &       $0.3807^{+0.0033}_{-0.0039}$ \\
$R_2$ ($\rsol$)           &       $0.2999^{+0.0044}_{-0.0043}$ &       $0.2938^{+0.0059}_{-0.0055}$ \\
\enddata

\end{deluxetable}

\subsubsection{Errors in the atmosphere models}

We have assumed limb darkening coefficients fixed to the theoretical
values from {\sc phoenix} model atmospheres throughout the analysis.
We note that the same atmosphere models have issues reproducing the
observed spectra of M-dwarfs in the optical, which raises the
possibility of systematic errors in the limb darkening law.  In the
literature, this is conventionally addressed by analyzing photometry
in multiple passbands, which is sensitive to differences in the limb
darkening as a function of wavelength.  We do not have multi-band
photometry, but note this could be an important area for future work.

It has long been recognized (e.g. \citealt{1972ApJ...174..617N} and
references therein) that changes in the radius of the eclipsed star
mimic changes in the limb darkening.  In a system with near-equal
stars such as the present, it is reasonable to expect any systematic
error in the limb darkening from the models should affect both stars
similarly, and thus to first order, the dominant effect will be on the
sum of the radii (and the inclination, but any uncertainties here have
negligible effect on the final parameters).

We verified this using simulations where the square root limb
darkening coefficients were varied, finding a strong correlation
between $(R_1+R_2)/a$ and the integral of the specific intensity over
the stellar disc, which is $1-u/3-u'/5$ for the square root law (it is
also correlated to a lesser extent with $J$ and $\cos i$).  This
quantity is the normalization term in the eclipse depths in the
photometric model, so it is not surprising that it should directly
influence the quantities derived from the absolute eclipse depth.
This is predominantly a concern for the interpretation of the sum of
the radii: we find even quite large changes in the limb darkening law
do not significantly alter the individual radii compared to their
uncertainties in our adopted model.

\subsection{Third light}
\label{mod_l3_sect}

We now examine our assumption of no third light.  While the proper
motion evidence (see Figure \ref{third_light_images}) argues it is
unlikely there are any background stars contributing at a significant
level in the photometric aperture, the relatively low proper motion of
our target does not allow this possibility to be completely eliminated
given the limited angular resolution of the first epoch imaging data.
Common proper motion companions are also permitted at intermediate
separations below the angular resolution of the SDSS data, but still
in wide enough orbits or with mass ratios much less than unity, where
they would not be detected by the presence of additional lines in the
spectra or radial velocity drift over the approximately 1 year
baseline available.

Therefore, to examine any constraints on third light which can be
placed directly from the light curves, and the effect on the derived
parameters, we ran an additional set of Monte Carlo simulations using
the basic ``non-eclipsed spots'' model on the primary discussed in \S
\ref{spot-results-sect} and Table \ref{fit-spot-params}, allowing
$L_3$ to vary.  For simplicity, a uniform prior was assumed.  We find
$L_3 < 0.029$ at $95\%$ confidence.  As discussed by
\citet{1972ApJ...174..617N}, the dominant parameter mimic is between
third light and inclination, and our results confirm this, finding the
uncertainties in $\cos i$ and the central surface brightness ratio 
parameters were slightly inflated.  The other light curve parameters
and radii derived from these chains are indistinguishable within the
uncertainties from the results where $L_3$ was not varied.

\subsection{Summary and adopted system parameters}
\label{mod_sum_sect}

Of the sources of uncertainty we have considered, spots dominate.
Since the stars hosting the spots and the spot configuration are
mostly unknown, we adopt the union of the six spotted solutions
discussed in \S \ref{add_spot_sect}, giving them equal weight.  The
final parameter estimates and their uncertainties are reported in
Table \ref{adopted-params}, derived from the posterior samples
produced by the Monte Carlo simulations, adopting the median and
$68.3\%$ central confidence intervals as the central value and
uncertainty (see \S \ref{mod_sol_sect}).

\begin{deluxetable}{lrr}
\tabletypesize{\normalsize}
\tablecaption{\label{adopted-params} Adopted physical parameters and
  uncertainties, combining six spot configurations.}
\tablecolumns{3}

\tablehead{
\colhead{Parameter} & \multicolumn{2}{c}{Value} \\
 & \colhead{C08\tablenotemark{a}} & \colhead{L96\tablenotemark{a}}
}

\startdata
$M_1$ ($\msol$)           &\multicolumn{2}{c}{$0.3946 \pm 0.0023$} \\
$M_2$ ($\msol$)           &\multicolumn{2}{c}{$0.2745 \pm 0.0012$} \\
$(R_1+R_2)$ ($\rsol$)     &\multicolumn{2}{c}{$0.6834^{+0.0046}_{-0.0028}$} \\
$R_1$ ($\rsol$)           &\multicolumn{2}{c}{$0.3860^{+0.0055}_{-0.0052}$} \\
$R_2$ ($\rsol$)           &\multicolumn{2}{c}{$0.2978^{+0.0049}_{-0.0046}$} \\
 \\
$\log g_1$                &\multicolumn{2}{c}{$4.861 \pm 0.012$} \\
$\log g_2$                &\multicolumn{2}{c}{$4.929 \pm 0.014$} \\
\\
$T_{\rm eff,1}$\tablenotemark{a} (K) & $3061 \pm 162$     &$3191 \pm 164$ \\
$T_{\rm eff,2}$\tablenotemark{a} (K) & $2952 \pm 163$     &$3079 \pm 166$ \\
$T_{\rm eff,2}/T_{\rm eff,1}$ & $0.959^{+0.049}_{-0.025}$ &$0.960^{+0.048}_{-0.025}$ \\
\\
$L_{\rm bol,1}$ ($\lsol$)     & $0.0119 \pm 0.0025$       &$0.0141 \pm 0.0028$ \\
$L_{\rm bol,2}$ ($\lsol$)     & $0.0061 \pm 0.0014$       &$0.0073 \pm 0.0016$ \\
\\
$(m-M)$                       & $3.40 \pm 0.22$           &$3.53 \pm 0.22$ \\
$d$ (pc)                      & $48.1 \pm 5.0$            &$51.0 \pm 5.2$ \\
\\
$U$\tablenotemark{b} (km/s)   &$ 56.1 \pm 3.1$            &$ 57.6 \pm 3.2$ \\
$V$ (km/s)                    &$-44.4 \pm 1.7$            &$-44.8 \pm 1.8$ \\
$W$ (km/s)                    &$-10.8 \pm 3.2$            &$ -9.1 \pm 3.4$ \\
\enddata

\tablenotetext{a}{Effective temperatures and bolometric corrections
  used. C08: \citet{2008MNRAS.389..585C}; L96:
  \citet{1996ApJS..104..117L}.  A $150\ {\rm K}$ systematic
  uncertainty was assumed for the effective temperatures, and the
  color equations given in the 2MASS explanatory supplement were used
  to convert the observed $JHK$ photometry to the CIT system when
  using the L96 tabulation.}
\tablenotetext{b}{Adopting the definition of positive $U$ toward
  the Galactic center.  Calculated using the method of
  \citet{1987AJ.....93..864J}.}

\end{deluxetable}

We compute UVW space motions for \theobject\ using the
method of \citet{1987AJ.....93..864J}, with the distance from the
eclipsing binary solution, $\gamma$ velocity from Table
\ref{fit-spot-params}, and position and proper motions from 
Table \ref{photparams}.   We find this object is in the old Galactic
disk population, following the method and definitions in
\citet{1992ApJS...82..351L}.

For very precise work, even matters as seemingly trivial as physical
constants can be important, so we briefly discuss our assumptions in
this regard.  We adopt the 2009 IAU values of $G \msol$ and the
astronomical unit (see the Astronomical Almanac 2011), and follow
\citet{2000asqu.book.....C} in adopting the value of the solar
photospheric radius from \citet{1998ApJ...500L.195B}\footnote{A useful
  compilation of solar data may be found in ``Basic Astronomical Data
  for the Sun'', by Eric Mamajek (University of Rochester, NY, USA),
  available at {\tt
    http://www.pas.rochester.edu/\~{}emamajek/sun.txt}.}.  For the
solar effective temperature, we use the value of $5781\ {\rm K}$ from
\citet{1998A&A...333..231B}, which is based on measurements of the
solar constant by \citet{1982SoEn...28..385D}.

The final constant, the solar bolometric absolute magnitude, is a more
thorny issue, as discussed by \citet{1998A&A...333..231B}, and
extensively by \citet{2010AJ....140.1158T}.  When using tables of
bolometric corrections, it is vital to adopt the consistent value of
this quantity in accordance with the table.  We therefore use the
appropriate values in each case in the following section.

\subsection{Effective temperatures, bolometric luminosities and distance}
\label{mod_teff_sect}

In order to estimate effective temperatures (and thus, bolometric
luminosities, and the distance), an external estimate of the effective
temperature of one of the binary components is required, in addition
to bolometric corrections.  For M-dwarfs, there are large systematic
uncertainties in the effective temperature scale
(e.g. \citealt{1997ARA&A..35..137A,1998ApJ...497..354L}), with
disagreement at the few hundred degrees Kelvin level among different
authors.  Many of the early difficulties were due to the lack of
model atmospheres (which are usually needed to integrate the full SED
from the available measurements to obtain $L_{\rm bol}$), and
significant improvement on this front was made in the 1990s.  The
availability of a much larger sample of angular diameter measurements
from interferometry should further improve the situation as this
provides a much more direct method to estimate $\teff$ for single
stars, but there are relatively few temperature scales available at
the present time using this information.

We show results from two different inversions to illustrate the
typical range of parameters.  Both scales we use were derived with
model atmospheres rather than blackbodies, and in both cases we use
the measured $\icous-J_{\rm 2MASS}$, $\icous-H_{\rm 2MASS}$ and
$\icous-K_{\rm 2MASS}$ colors from Table \ref{photparams} in
conjunction with the component radii and $\icous$-band light ratio
from the adopted eclipsing binary solution.  The results from these
different pairings of bandpasses were found to be consistent within
the uncertainties, so we report the union of the results.

The two sets of effective temperature and bolometric corrections
adopted were from \citet{2008MNRAS.389..585C}, which is a recent
determination using interferometric angular diameter measurements to
derive $T_{\rm eff}$, and from \citet{1996ApJS..104..117L}, which is
the basis for several recent works and tabulations of the effective
temperature scale, including \citet{1998ApJ...497..354L},
\citet{1999ApJ...525..466L}, and \citet{2007AJ....134.2340K}.

For the \citet{1996ApJS..104..117L} scale, we fit polynomials to their
tabulated data, omitting three objects these authors found to be metal
poor from our fits: Gl~129, LHS~343, and LHS~377.  We follow
\citet{1996ApJS..104..117L} in adopting a systematic uncertainty of
$150\ {\rm K}$, which appears to be consistent with the differences we
find between the two determinations.

Our results for \theobject\ are reported in Table \ref{adopted-params}.

\section{Discussion}

\begin{figure}
\centering
\includegraphics[angle=270,width=3.2in]{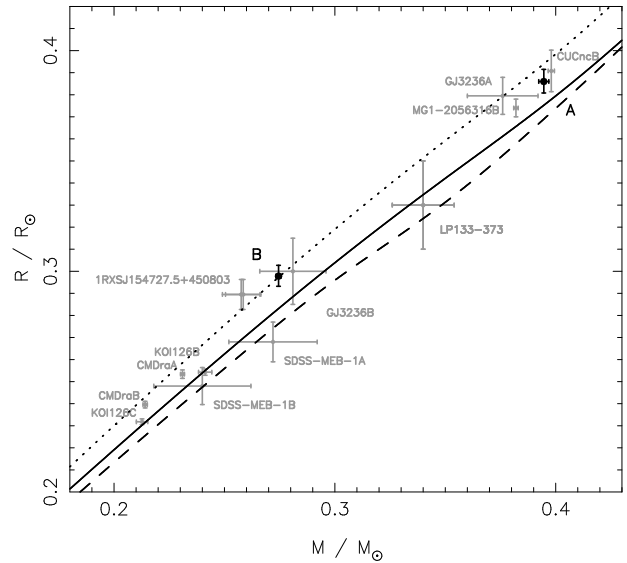}
\caption{Mass-radius relation for components of detached,
  double-lined, double-eclipsing binary systems below $0.4\ \msol$.
  \theobject\ is shown as bold, black points, and the literature
  systems from Table \ref{lit_eb_pars} are shown in gray, excluding
  CCDM~J04404+3127B,C, which does not have a full solution available
  at present, but including KOI~126.  The lines show stellar evolution
  models from \citet{1998A&A...337..403B} for $10\ {\rm Gyr}$ age,
  $[{\rm M/H}] = 0$ (solid line), $[{\rm M/H}] = -0.5$ (dashed line),
  and the $[{\rm M/H}] = 0$ model with the radius inflated by $5\%$,
  corresponding to $\rho = 1.05$ (dotted line).}
\label{mrrel}
\end{figure}

\begin{figure}
\centering
\includegraphics[angle=270,width=3.2in]{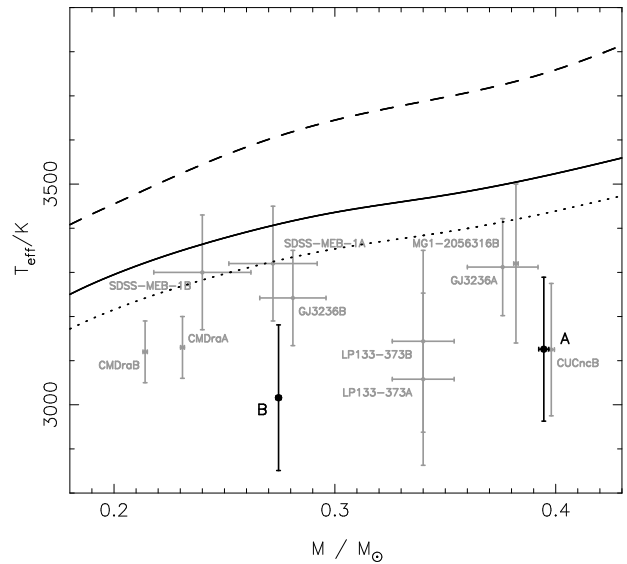}
\caption{Mass-effective temperature relation for components of detached,
  double-lined, double-eclipsing binary systems below $0.4\ \msol$.
  Lines and points as Figure \ref{mrrel}.  KOI~126B,C and
  1RXS~J154727.5+450803 do not have reported effective temperatures,
  so they do not appear on this diagram.  The dotted line shows the
  prediction from assuming $\rho = 1.05$ and that the bolometric
  luminosity is preserved.}
\label{mtrel}
\end{figure}

We now compare our results for \theobject\ with theoretical models
from \citet{1998A&A...337..403B} and other literature objects with
shorter periods summarized in Table \ref{lit_eb_pars}.  We first show
the conventional mass-radius diagram in Figure \ref{mrrel}, and then
the corresponding mass-effective temperature ($\teff$) diagram in
Figure \ref{mtrel}.

\begin{deluxetable*}{lrllllc}
\tabletypesize{\normalsize}
\tablecaption{\label{lit_eb_pars} Detached, double-lined,
  double-eclipsing main sequence EB components below
  $0.4\ \msol$\tablenotemark{a}}
\tablecolumns{7}

\tablehead{
\colhead{Name} & \colhead{Period} & \colhead{$M$}     & \colhead{$R$}     & \colhead{$\teff$}  & \colhead{$[{\rm M/H}]$}    & \colhead{Source\tablenotemark{b}} \\
               & \colhead{(days)} & \colhead{($\msol$)} & \colhead{($\rsol$)} & \colhead{(K)} & & 
}

\startdata
SDSS-MEB-1~A                          & $0.407$ &$0.272 \pm 0.020$    &$0.268 \pm 0.0090$          &$3320 \pm 130$  &\ldots           & 1 \\
SDSS-MEB-1~B                          &         &$0.240 \pm 0.022$    &$0.248 \pm 0.0084$          &$3300 \pm 130$  &                 & \\
GJ~3236~A\tablenotemark{c}            & $0.771$ &$0.376 \pm 0.016$    &$0.3795 \pm 0.0084$         &$3312 \pm 110$  &\ldots           & 2 \\
GJ~3236~B\tablenotemark{c}            &         &$0.281 \pm 0.015$    &$0.300 \pm 0.015$           &$3242 \pm 108$  &                 & \\
CM~Dra~A                              & $1.27$  &$0.2310 \pm 0.0009$  &$0.2534 \pm 0.0019$         &$3130 \pm 70$   &$[-1,-0.6]$      & 3 \\
CM~Dra~B                              &         &$0.2141 \pm 0.0010$  &$0.2396 \pm 0.0015$         &$3120 \pm 70$   &                 & \\
LP~133-373~A                          & $1.63$  &$0.340 \pm 0.014$    &$0.33 \pm 0.02$             &$3058 \pm 195$  &\ldots           & 4 \\
LP~133-373~B                          &         &same                 &same                        &$3144 \pm 206$  &\ldots           &   \\
MG1-2056316~B                         & $1.72$  &$0.382 \pm 0.001$    &$0.374 \pm 0.002 \pm 0.002$ &$3320 \pm 180$  &\ldots           & 5 \\
KOI~126~B\tablenotemark{d}            & $1.77$  &$0.2413 \pm 0.0030$  &$0.2543 \pm 0.0014$         &\ldots          &$+0.15 \pm 0.08$ & 6 \\
KOI~126~C\tablenotemark{d}            &         &$0.2127 \pm 0.0026$  &$0.2318 \pm 0.0013$         &\ldots          &                 & \\
CCDM~J04404+3127~B,C\tablenotemark{e} & $2.05$  &\ldots               &\ldots                      &\ldots          &\ldots           & 7 \\
CU~Cnc~B                              & $2.77$  &$0.3980 \pm 0.0014$  &$0.3908 \pm 0.0094$         &$3125 \pm 150$  &\ldots           & 8 \\
1RXS~J154727.5+450803~A               & $3.55$  &$0.2576 \pm 0.0085$  &$0.2895 \pm 0.0068$         &\ldots          &\ldots           & 9 \\
1RXS~J154727.5+450803~B               &         &$0.2585 \pm 0.0080$  &same                        &\ldots          &                 & \\
\enddata

\tablenotetext{a}{This mass criterion has been applied to keep the
  number of objects in the table and plots manageable, and is not
  intended to necessarily be physically meaningful beyond being
  appropriate for comparison to the present system.}
\tablenotetext{b}{(1) \citet{2008ApJ...684..635B}, (2)
  \citet{2009ApJ...701.1436I}, (3) \citet{2009ApJ...691.1400M}, (4)
  \citet{2007ApJ...661.1112V}, (5) \citet{2011ApJ...728...48K}, (6)
  \citet{2011Sci...331..562C}, (7) \citet{2010ApJ...716.1522S}, (8)
  \citet{2003A&A...398..239R,1999A&A...350L..39D}, (9)
  \citet{2011AJ....141..166H}.}
\tablenotetext{c}{Parameters determined giving equal weight to all
  three models, following \citet{2011AJ....141..166H}.}
\tablenotetext{d}{While not double-lined, this object is a special
  case as it is still possible to solve for the masses and radii of
  both M-dwarfs independent of M-dwarf models.  The period given is
  that for the inner M-dwarf binary as this is presumably the
  appropriate one for estimation of activity levels.}
\tablenotetext{e}{Full solution not available.}

\end{deluxetable*}

We also perform a quantitative comparison with the same models.  We
follow \citet{2007ApJ...671L..65T} in defining a parameter equal to
the ratio of the observed radius to that predicted by the models given
the observed mass, although we denote this parameter by $\rho$
instead, to avoid any potential confusion with the $\beta$ parameter
defined by \citet{2007A&A...472L..17C} and used by
\citet{2010ApJ...718..502M}, which is not the same.

As argued in \citet{2007ApJ...671L..65T}, there is evidence that the
bolometric luminosities from the models are not seriously in error.
If this is the case, a corresponding factor of $\rho^{-1/2}$ must be
applied to the effective temperatures from the models when inflating
the radii, in order to preserve the bolometric luminosity.  We show
results for this scenario in the tables and plots.  Note, however,
that the bolometric luminosity is not preserved in several recent
theoretical works
\citep{2007A&A...472L..17C,2010ApJ...718..502M,2011arXiv1106.1452M},
so the temperatures derived from this procedure should not be
considered definitive.

As is often the case in eclipsing binaries, while the individual
masses of the objects are well-determined from the radial velocities,
the sum of the radii is much better determined than the individual
radii, modulo limb darkening uncertainties as discussed in
\S \ref{mod_ld_sect}.  We therefore compute, in addition to the $\rho$
values for the individual stars, a ``composite'' value comparing the
model-predicted radius sum given the individual observed masses, to
the measured radius sum.  This quantity, which we call $\rho_{1+2}$,
has much smaller formal uncertainties than the individual $\rho_1$ and
$\rho_2$  values, and measures essentially a ``weighted average
inflation'' for the whole system.  These quantities, and the
corresponding effective temperatures predicted from preserving the
bolometric luminosity, are given for a range of assumptions for the
age and metallicity in Table \ref{rho_table}.  While we caution
over-interpretation of the $\rho_{1+2}$ results due to limb darkening
uncertainty, the individual values should be much less prone to this
problem.

\begin{deluxetable*}{rrrrrrr}
\tabletypesize{\normalsize}
\tablecaption{\label{rho_table} Quantitative mass-radius comparison
  of \theobject\ with models.}
\tablecolumns{7}

\tablehead{
\colhead{Age} & \colhead{$[{\rm M/H}]$} &
\colhead{$\rho_1$} & \colhead{$\rho_2$} & \colhead{$T_{\rm eff,1}$} &
\colhead{$T_{\rm eff,2}$} & \colhead{$T_{\rm eff,2}/T_{\rm eff,1}$} \\
\colhead{(Gyr)} & & \multicolumn{2}{c}{$\rho_{1+2}$} & \colhead{(K)} & \colhead{(K)} & 
}

\startdata
$ 1$ & $ 0.0$ & $1.054 \pm 0.016$ & $1.096 \pm 0.018$ & $3426 \pm 27$ & $3255 \pm 26$ & $0.950 \pm 0.013$ \\
$ 1$ & $-0.5$ & $1.070 \pm 0.016$ & $1.130 \pm 0.018$ & $3621 \pm 28$ & $3387 \pm 27$ & $0.935 \pm 0.013$ \\
$ 5$ & $ 0.0$ & $1.037 \pm 0.015$ & $1.072 \pm 0.017$ & $3453 \pm 27$ & $3293 \pm 27$ & $0.954 \pm 0.013$ \\
$ 5$ & $-0.5$ & $1.055 \pm 0.016$ & $1.101 \pm 0.018$ & $3648 \pm 29$ & $3436 \pm 28$ & $0.942 \pm 0.013$ \\
$10$ & $ 0.0$ & $1.029 \pm 0.015$ & $1.052 \pm 0.017$ & $3468 \pm 27$ & $3324 \pm 27$ & $0.959 \pm 0.013$ \\
$10$ & $-0.5$ & $1.047 \pm 0.016$ & $1.078 \pm 0.017$ & $3666 \pm 29$ & $3475 \pm 28$ & $0.948 \pm 0.013$ \\
\\
\hline
\\
$ 1$ & $ 0.0$ & \multicolumn{2}{c}{$1.0706^{+0.0091}_{-0.0055}$} & $3396 \pm 13$ & $3291 \pm 12$ & $0.96906 \pm 0.00035$ \\
$ 1$ & $-0.5$ & \multicolumn{2}{c}{$1.0941^{+0.0092}_{-0.0055}$} & $3579 \pm 14$ & $3439 \pm 13$ & $0.96096 \pm 0.00045$ \\
$ 5$ & $ 0.0$ & \multicolumn{2}{c}{$1.0505^{+0.0089}_{-0.0053}$} & $3428 \pm 13$ & $3323 \pm 12$ & $0.96940 \pm 0.00038$ \\
$ 5$ & $-0.5$ & \multicolumn{2}{c}{$1.0735^{+0.0091}_{-0.0055}$} & $3614 \pm 14$ & $3477 \pm 13$ & $0.96195 \pm 0.00050$ \\
$10$ & $ 0.0$ & \multicolumn{2}{c}{$1.0378^{+0.0087}_{-0.0052}$} & $3451 \pm 13$ & $3344 \pm 12$ & $0.96915 \pm 0.00040$ \\
$10$ & $-0.5$ & \multicolumn{2}{c}{$1.0590^{+0.0091}_{-0.0056}$} & $3642 \pm 15$ & $3504 \pm 13$ & $0.96221 \pm 0.00052$ \\
\enddata

\end{deluxetable*}

\subsection{Mass-radius comparison}

We first limit our comparisons to the mass-radius plane.  If the
metallicity is solar, the sum of radii is inflated at the $7 \sigma$
level compared to the models, even for $10\ {\rm Gyr}$ age, which
produces the largest model radii.  The discrepancies are not
as significant in the individual radii, with the primary $2 \sigma$
and the secondary $3 \sigma$ larger than the model predictions.  We
also note that the ``eclipsed spots'' model on the secondary, which we
discarded in \S \ref{mod_spot_sect}, does not solve the radius
discrepancy, having a secondary radius (and sum of radii) even further
from the model predictions.

Youth is unlikely as an explanation given the old disk kinematics
of our target.  We  estimate an age of approximately $120\ {\rm Myr}$
at solar metallicity would be needed to bring the radius sum into
agreement with the models, yet such objects should also rotate
extremely rapidly and exhibit strong H$\alpha$ emission
(e.g. \citealt{2000AJ....119.1303T,2010MNRAS.408..475H}, and
references therein), neither of which are seen.

Metallicity may provide a viable explanation for the observed radii
given the present uncertainties (both in the observations and the
models, where for the latter it is possible the effect of metallicity
on radius is underpredicted; see \S \ref{intro_sect} and
\citealt{2007ApJ...660..732L}), although the object would most likely
need to be extremely metal rich.  This would place it in the tail of
the metallicity distribution for the old Galactic disk population,
which is likely to peak around $[{\rm M/H}] \approx -0.5$
(e.g. \citealt{1992ApJS...82..351L} and references therein).

Given that metallicity has difficulty reproducing radii in
other well-observed systems (see \S \ref{intro_sect}) and the
kinematic evidence, we proceed to examine other possible causes of
the inflated radii.

It is tempting to suggest that the inflation in \theobject\ results
from elevated activity levels in one or both components.  While the
lack of H$\alpha$ emission, unless this is rendered undetectable by
the limited signal-to-noise ratio of our spectra, may be a difficulty
for this scenario to explain, it should be noted that more modest
levels of activity can result in weakened H$\alpha$ emission, or even
absorption, but may still produce sufficient spot coverage to explain
the inflation.

This system straddles (in mass) the full convection limit, where a
substantial increase in the observed activity lifetimes for M-dwarfs
occurs (e.g. \citealt{2008AJ....135..785W}).  This would argue that
the secondary is more likely to be active and thus the source of the
inflation.  The secondary lines in the spectrum are also more
difficult to observe due to its lower luminosity, which would make an
H$\alpha$ line easier to hide in the noise.  Identification of the
star hosting the out of eclipse modulations may shed additional light
on which of the components might be responsible for the inflation.

As discussed by \citet{2010ApJ...718..502M}, the radius inflation
inferred from eclipsing binary analyses often implies extremely high
spot filling factors.  We estimate that $\rho = 1.05$ corresponds to
$\beta = 0.1$.  Assuming a spot temperature contrast of $T_{\rm
  eff,spot}/T_{\rm eff,phot} = 0.9$, this would require a filling
factor of approximately $30\%$.  This is much larger than needed to
produce the observed out of eclipse modulation, and would mean there
is a substantial longitudinally symmetric spot component, if the
inflation is indeed caused by spots.  We showed in \S
\ref{add_spot_sect} that such a level of spots does slightly modify
the component radii if they are not eclipsed, reducing $R_2$ and
increasing $R_1$.  In practice, this may slightly reduce the required
filling factor depending on where the spots are located, but unless
the spot coverage is much larger than we assumed, the systematic
errors in the radii resulting from such spots are still insufficient
to explain the observations without invoking inflation.

Finally, on suggestion of the referee, we note that this object may
indicate the need to revisit the equation of state for very
low mass stars, which was discussed as a possibility to explain the
radius inflation before the activity hypothesis gained prevalence
(e.g. \citealt{2004PhDT........12L,2002ApJ...567.1140T}).  The
equation of state was a significant source of difficulty in early work
attempting to model very low mass stars \citep{1995ApJ...451L..29C},
and although substantial progress on this front has been made
(see the reviews by
\citealt{2000ARA&A..38..337C,2005astro.ph..9798C}), it is still an
open question.

\subsection{Mass-$T_{\rm eff}$ comparison}

Comparisons in only the mass-radius plane ignore important information
contained in the effective temperatures.  The correct physical
model must explain all of the observations simultaneously, so we
now proceed to examine the temperatures.  This comparison is more
problematic than in mass-radius for a number of reasons.  The main
observational issues are the difficulty of constraining effective
temperatures (see \S \ref{mod_teff_sect}) and metallicities for the
eclipsing binaries.

As noted in \S \ref{intro_sect}, unlike the radius, the effective
temperature predicted from models does depend quite strongly on
metallicity, because the bolometric luminosity is a function of
metallicity.  As the metallicity is decreased, the model bolometric
luminosity and effective temperature increase, and the radius 
decreases by a small amount.  While metallicity does complicate the
interpretation of the mass-$\teff$ diagram, this a key argument
against metallicity as the explanation for all of the inflated radii
in the eclipsing binary sample, as discussed already in \S
\ref{intro_sect}.  In this regard, CM~Dra is puzzling as several
authors have claimed this object is metal poor, which would further
exacerbate the effective temperature discrepancy with the models.

Given the present uncertainties, it is not clear if the effective
temperature difference between \theobject\ and the models is
significant.  However, it does lie at the low-temperature end of the
eclipsing binary results, and this lends support to the hypothesis
that this object is more metal rich than the other eclipsing
binaries.

\subsection{Future work}

It is clear that \theobject\ offers one of the best prospects to
observationally test the causes of inflated radii in eclipsing
binaries, but further observations are needed.  The most fruitful
avenues would be to pursue a determination of the metallicity,
improvement of the system parameters with a particular focus to
investigating potential systematics in the radii (e.g. due to spots),
and constraining the activity levels in the components by independent
means (for example, X-ray emission, or the Ca{\sc ii} H and K lines).

While we have mentioned the non-synchronized spin and orbit, we were
unable to take full advantage of this property with the available
observational data.  As shown in \S \ref{mod_spot_sect}, 
the effect of spots should be larger on the primary eclipse, so we
advocate an intensive campaign on these events, preferably to
obtain multiple, complete eclipses with the same telescope and
detector system at a range of rotational phases and in multiple
wavelengths.  The primary eclipses are observable from Europe and
Scandinavia during the winter 2011-2012 season, where excellent
observational facilities are available.  We remind observers that
good-quality out of eclipse monitoring will be important for the
interpretation of the eclipses themselves, and the spot distributions
are likely to evolve (indeed, Figure \ref{ooe-fit} indicates that we
may have already seen this in the existing data), so the monitoring
must be contemporaneous, and preferably also taken at multiple
wavelengths to constrain the spot temperatures.

\acknowledgments The MEarth team gratefully acknowledges funding from
the David and Lucile Packard Fellowship for Science and Engineering
(awarded to DC) and the National Science Foundation (NSF) under grant
number AST-0807690.  SNQ, DWL, and GAE acknowledge partial support
from the NASA Kepler mission under cooperative agreement NCC2-1390,
and GT acknowledges partial support from NSF grant AST-1007992.  We
thank Allyson Bieryla for assistance with the Clay telescope
observations, David Kipping for discussions of appropriate priors for
use in Monte Carlo simulations, Eric Mamajek for compiling and
maintaining the ``Basic Astronomical Data for the Sun'' page, Todd
Vaccaro for clarification of the parameters of LP~133-373, Andrew West
for discussions of M-dwarf activity, and Edo Berger and the 2011
Harvard Astronomy 100 undergraduate class for assistance with the FAST
observations.  The anonymous referee is thanked for a prompt and
helpful report that improved the manuscript.  The MEarth team is
greatly indebted to the staff at the Fred Lawrence Whipple Observatory
for their efforts in construction and maintenance of the facility, and
would like to thank Wayne Peters, Ted Groner, Karen Erdman-Myres,
Grace Alegria, Rodger Harris, Bob Hutchins, Dave Martina, Dennis
Jankovsky and Tom Welsh for their support.

This research has made extensive use of data products from the Two
Micron All Sky Survey, which is a joint project of the University of
Massachusetts and the Infrared Processing and Analysis Center /
California Institute of Technology, funded by NASA and the NSF, NASA's
Astrophysics Data System (ADS) bibliographic services, and the SIMBAD
database, operated at CDS, Strasbourg, France.  The Digitized Sky
Surveys were produced at the Space Telescope Science Institute under
U.S. Government grant NAG W-2166. The images of these surveys are
based on photographic data obtained using the Oschin Schmidt Telescope
on Palomar Mountain and the UK Schmidt Telescope. The plates were
processed into the present compressed digital form with the permission
of these institutions.

Funding for the SDSS and SDSS-II has been provided by the Alfred
P. Sloan Foundation, the Participating Institutions, the National
Science Foundation, the U.S. Department of Energy, the National
Aeronautics and Space Administration, the Japanese Monbukagakusho, the
Max Planck Society, and the Higher Education Funding Council for
England. The SDSS Web Site is {\tt http://www.sdss.org/}.
The SDSS is managed by the Astrophysical Research Consortium for the
Participating Institutions. The Participating Institutions are the
American Museum of Natural History, Astrophysical Institute Potsdam,
University of Basel, University of Cambridge, Case Western Reserve
University, University of Chicago, Drexel University, Fermilab, the
Institute for Advanced Study, the Japan Participation Group, Johns
Hopkins University, the Joint Institute for Nuclear Astrophysics, the
Kavli Institute for Particle Astrophysics and Cosmology, the Korean
Scientist Group, the Chinese Academy of Sciences (LAMOST), Los Alamos
National Laboratory, the Max-Planck-Institute for Astronomy (MPIA),
the Max-Planck-Institute for Astrophysics (MPA), New Mexico State
University, Ohio State University, University of Pittsburgh,
University of Portsmouth, Princeton University, the United States
Naval Observatory, and the University of Washington.

{\it Facilities:} \facility{FLWO:1.5m (FAST, TRES)}

\bibliography{all}

\begin{thebibliography}{108}
\expandafter\ifx\csname natexlab\endcsname\relax\def\natexlab#1{#1}\fi

\bibitem[{{Abazajian} {et~al.}(2009){Abazajian}, {Adelman-McCarthy},
  {Ag{\"u}eros}, {Allam}, {Allende Prieto}, {An}, {Anderson}, {Anderson},
  {Annis}, {Bahcall}, \& et~al.}]{2009ApJS..182..543A}
{Abazajian}, K.~N., {et~al.} 2009, \apjs, 182, 543

\bibitem[{{Allard} {et~al.}(1997){Allard}, {Hauschildt}, {Alexander}, \&
  {Starrfield}}]{1997ARA&A..35..137A}
{Allard}, F., {Hauschildt}, P.~H., {Alexander}, D.~R., \& {Starrfield}, S.
  1997, \araa, 35, 137

\bibitem[{{Andersen}(1991)}]{1991A&ARv...3...91A}
{Andersen}, J. 1991, \aapr, 3, 91

\bibitem[{{Baraffe} {et~al.}(1998){Baraffe}, {Chabrier}, {Allard}, \&
  {Hauschildt}}]{1998A&A...337..403B}
{Baraffe}, I., {Chabrier}, G., {Allard}, F., \& {Hauschildt}, P.~H. 1998, \aap,
  337, 403

\bibitem[{{Barnard}(1916)}]{1916AJ.....29..181B}
{Barnard}, E.~E. 1916, \aj, 29, 181

\bibitem[{{Barnes} \& {Collier Cameron}(2001)}]{2001MNRAS.326..950B}
{Barnes}, J.~R., \& {Collier Cameron}, A. 2001, \mnras, 326, 950

\bibitem[{{Barnes} {et~al.}(2004){Barnes}, {James}, \& {Collier
  Cameron}}]{2004MNRAS.352..589B}
{Barnes}, J.~R., {James}, D.~J., \& {Collier Cameron}, A. 2004, \mnras, 352,
  589

\bibitem[{{Barnes} {et~al.}(2011){Barnes}, {Jeffers}, \&
  {Jones}}]{2011MNRAS.412.1599B}
{Barnes}, J.~R., {Jeffers}, S.~V., \& {Jones}, H.~R.~A. 2011, \mnras, 412, 1599

\bibitem[{{Bean} {et~al.}(2006){Bean}, {Sneden}, {Hauschildt}, {Johns-Krull},
  \& {Benedict}}]{2006ApJ...652.1604B}
{Bean}, J.~L., {Sneden}, C., {Hauschildt}, P.~H., {Johns-Krull}, C.~M., \&
  {Benedict}, G.~F. 2006, \apj, 652, 1604

\bibitem[{{Benedict} {et~al.}(1998){Benedict}, {McArthur}, {Nelan}, {Story},
  {Whipple}, {Shelus}, {Jefferys}, {Hemenway}, {Franz}, {Wasserman},
  {Duncombe}, {van Altena}, \& {Fredrick}}]{1998AJ....116..429B}
{Benedict}, G.~F., {et~al.} 1998, \aj, 116, 429

\bibitem[{{Berger} {et~al.}(2006){Berger}, {Gies}, {McAlister}, {ten
  Brummelaar}, {Henry}, {Sturmann}, {Sturmann}, {Turner}, {Ridgway},
  {Aufdenberg}, \& {M{\'e}rand}}]{2006ApJ...644..475B}
{Berger}, D.~H., {et~al.} 2006, \apj, 644, 475

\bibitem[{{Berta} {et~al.}(2011){Berta}, {Charbonneau}, {Bean}, {Irwin},
  {Burke}, {D{\'e}sert}, {Nutzman}, \& {Falco}}]{2011ApJ...736...12B}
{Berta}, Z.~K., {Charbonneau}, D., {Bean}, J., {Irwin}, J., {Burke}, C.~J.,
  {D{\'e}sert}, J.-M., {Nutzman}, P., \& {Falco}, E.~E. 2011, \apj, 736, 12

\bibitem[{{Bessel}(1990)}]{1990A&AS...83..357B}
{Bessel}, M.~S. 1990, \aaps, 83, 357

\bibitem[{{Bessell}(1990)}]{1990PASP..102.1181B}
{Bessell}, M.~S. 1990, \pasp, 102, 1181

\bibitem[{{Bessell} {et~al.}(1998){Bessell}, {Castelli}, \&
  {Plez}}]{1998A&A...333..231B}
{Bessell}, M.~S., {Castelli}, F., \& {Plez}, B. 1998, \aap, 333, 231

\bibitem[{{Blake} {et~al.}(2008){Blake}, {Torres}, {Bloom}, \&
  {Gaudi}}]{2008ApJ...684..635B}
{Blake}, C.~H., {Torres}, G., {Bloom}, J.~S., \& {Gaudi}, B.~S. 2008, \apj,
  684, 635

\bibitem[{{Bonfils} {et~al.}(2005){Bonfils}, {Delfosse}, {Udry}, {Santos},
  {Forveille}, \& {S{\'e}gransan}}]{2005A&A...442..635B}
{Bonfils}, X., {Delfosse}, X., {Udry}, S., {Santos}, N.~C., {Forveille}, T., \&
  {S{\'e}gransan}, D. 2005, \aap, 442, 635

\bibitem[{{Boyajian} {et~al.}(2011){Boyajian}, {von Braun}, {van Belle}, {ten
  Brummelaar}, {Ciardi}, {Henry}, {Lopez-Morales}, {McAlister}, {Ridgway},
  {Farrington}, {Goldfinger}, {Sturmann}, {Sturmann}, \&
  {Turner}}]{2010arXiv1012.0542B}
{Boyajian}, T.~S., {et~al.} 2011, in Cool Stars, Stellar Systems and the Sun
  XVI, ed. {C.~Johns-Krull, A.~West, \& M.~Browning}, Astronomical Society of
  the Pacific Conference Series, {in press (arXiv:1012.0542)}

\bibitem[{{Brown} \& {Christensen-Dalsgaard}(1998)}]{1998ApJ...500L.195B}
{Brown}, T.~M., \& {Christensen-Dalsgaard}, J. 1998, \apjl, 500, L195+

\bibitem[{{Buchhave} {et~al.}(2010){Buchhave}, {Bakos}, {Hartman}, {Torres},
  {Kov{\'a}cs}, {Latham}, {Noyes}, {Esquerdo}, {Everett}, {Howard}, {Marcy},
  {Fischer}, {Johnson}, {Andersen}, {F{\H u}r{\'e}sz}, {Perumpilly},
  {Sasselov}, {Stefanik}, {B{\'e}ky}, {L{\'a}z{\'a}r}, {Papp}, \&
  {S{\'a}ri}}]{2010ApJ...720.1118B}
{Buchhave}, L.~A., {et~al.} 2010, \apj, 720, 1118

\bibitem[{{Carney} {et~al.}(1989){Carney}, {Latham}, \&
  {Laird}}]{1989AJ.....97..423C}
{Carney}, B.~W., {Latham}, D.~W., \& {Laird}, J.~B. 1989, \aj, 97, 423

\bibitem[{{Carter} {et~al.}(2011){Carter}, {Fabrycky}, {Ragozzine}, {Holman},
  {Quinn}, {Latham}, {Buchhave}, {Van Cleve}, {Cochran}, {Cote}, {Endl},
  {Ford}, {Haas}, {Jenkins}, {Koch}, {Li}, {Lissauer}, {MacQueen}, {Middour},
  {Orosz}, {Rowe}, {Steffen}, \& {Welsh}}]{2011Sci...331..562C}
{Carter}, J.~A., {et~al.} 2011, Science, 331, 562

\bibitem[{{Casagrande} {et~al.}(2008){Casagrande}, {Flynn}, \&
  {Bessell}}]{2008MNRAS.389..585C}
{Casagrande}, L., {Flynn}, C., \& {Bessell}, M. 2008, \mnras, 389, 585

\bibitem[{{Chabrier} \& {Baraffe}(1995)}]{1995ApJ...451L..29C}
{Chabrier}, G., \& {Baraffe}, I. 1995, \apjl, 451, L29+

\bibitem[{{Chabrier} \& {Baraffe}(1997)}]{1997A&A...327.1039C}
---. 1997, \aap, 327, 1039

\bibitem[{{Chabrier} \& {Baraffe}(2000)}]{2000ARA&A..38..337C}
---. 2000, \araa, 38, 337

\bibitem[{{Chabrier} {et~al.}(2005){Chabrier}, {Baraffe}, {Allard}, \&
  {Hauschildt}}]{2005astro.ph..9798C}
{Chabrier}, G., {Baraffe}, I., {Allard}, F., \& {Hauschildt}, P.~H. 2005, in
  Resolved Stellar Populations, ed. {D.~Valls-Gabaud, \& M.~Chavez},
  Astronomical Society of the Pacific Conference Series, {in press
  (astro-ph/0509798)}

\bibitem[{{Chabrier} {et~al.}(2007){Chabrier}, {Gallardo}, \&
  {Baraffe}}]{2007A&A...472L..17C}
{Chabrier}, G., {Gallardo}, J., \& {Baraffe}, I. 2007, \aap, 472, L17

\bibitem[{{Claret}(2000)}]{2000A&A...363.1081C}
{Claret}, A. 2000, \aap, 363, 1081

\bibitem[{{Claret}(2004)}]{2004A&A...428.1001C}
---. 2004, \aap, 428, 1001

\bibitem[{{Coughlin} {et~al.}(2011){Coughlin}, {L{\'o}pez-Morales}, {Harrison},
  {Ule}, \& {Hoffman}}]{2011AJ....141...78C}
{Coughlin}, J.~L., {L{\'o}pez-Morales}, M., {Harrison}, T.~E., {Ule}, N., \&
  {Hoffman}, D.~I. 2011, \aj, 141, 78

\bibitem[{{Covey} {et~al.}(2007){Covey}, {Ivezi{\'c}}, {Schlegel},
  {Finkbeiner}, {Padmanabhan}, {Lupton}, {Ag{\"u}eros}, {Bochanski}, {Hawley},
  {West}, {Seth}, {Kimball}, {Gogarten}, {Claire}, {Haggard}, {Kaib},
  {Schneider}, \& {Sesar}}]{2007AJ....134.2398C}
{Covey}, K.~R., {et~al.} 2007, \aj, 134, 2398

\bibitem[{{Cox}(2000)}]{2000asqu.book.....C}
{Cox}, A.~N., ed. 2000, {Allen's astrophysical quantities}, 4th edn. (New York:
  AIP Press; Berlin: Springer)

\bibitem[{{Delfosse} {et~al.}(2000){Delfosse}, {Forveille}, {S{\'e}gransan},
  {Beuzit}, {Udry}, {Perrier}, \& {Mayor}}]{2000A&A...364..217D}
{Delfosse}, X., {Forveille}, T., {S{\'e}gransan}, D., {Beuzit}, J.-L., {Udry},
  S., {Perrier}, C., \& {Mayor}, M. 2000, \aap, 364, 217

\bibitem[{{Delfosse} {et~al.}(1999){Delfosse}, {Forveille}, {Udry}, {Beuzit},
  {Mayor}, \& {Perrier}}]{1999A&A...350L..39D}
{Delfosse}, X., {Forveille}, T., {Udry}, S., {Beuzit}, J., {Mayor}, M., \&
  {Perrier}, C. 1999, \aap, 350, L39

\bibitem[{{Demory} {et~al.}(2009){Demory}, {S{\'e}gransan}, {Forveille},
  {Queloz}, {Beuzit}, {Delfosse}, {di Folco}, {Kervella}, {Le Bouquin},
  {Perrier}, {Benisty}, {Duvert}, {Hofmann}, {Lopez}, \&
  {Petrov}}]{2009A&A...505..205D}
{Demory}, B.-O., {et~al.} 2009, \aap, 505, 205

\bibitem[{{Devor} {et~al.}(2008){Devor}, {Charbonneau}, {Torres}, {Blake},
  {White}, {Rabus}, {O'Donovan}, {Mandushev}, {Bakos}, {F{\H u}r{\'e}sz}, \&
  {Szentgyorgyi}}]{2008ApJ...687.1253D}
{Devor}, J., {et~al.} 2008, \apj, 687, 1253

\bibitem[{{Diaz-Cordoves} \& {Gimenez}(1992)}]{1992A&A...259..227D}
{Diaz-Cordoves}, J., \& {Gimenez}, A. 1992, \aap, 259, 227

\bibitem[{{Doi} {et~al.}(2010){Doi}, {Tanaka}, {Fukugita}, {Gunn}, {Yasuda},
  {Ivezi{\'c}}, {Brinkmann}, {de Haars}, {Kleinman}, {Krzesinski}, \& {French
  Leger}}]{2010AJ....139.1628D}
{Doi}, M., {et~al.} 2010, \aj, 139, 1628

\bibitem[{{Duncan} {et~al.}(1982){Duncan}, {Willson}, {Kendall}, {Harrison}, \&
  {Hickey}}]{1982SoEn...28..385D}
{Duncan}, C.~H., {Willson}, R.~C., {Kendall}, J.~M., {Harrison}, R.~G., \&
  {Hickey}, J.~R. 1982, Solar Energy, 28, 385

\bibitem[{{Eggen} \& {Sandage}(1967)}]{1967ApJ...148..911E}
{Eggen}, O.~J., \& {Sandage}, A. 1967, \apj, 148, 911

\bibitem[{{Fabricant} {et~al.}(1998){Fabricant}, {Cheimets}, {Caldwell}, \&
  {Geary}}]{1998PASP..110...79F}
{Fabricant}, D., {Cheimets}, P., {Caldwell}, N., \& {Geary}, J. 1998, \pasp,
  110, 79

\bibitem[{{Ford}(2005)}]{2005AJ....129.1706F}
{Ford}, E.~B. 2005, \aj, 129, 1706

\bibitem[{{Ford}(2006)}]{2006ApJ...642..505F}
---. 2006, \apj, 642, 505

\bibitem[{{Gizis}(1997)}]{1997AJ....113..806G}
{Gizis}, J.~E. 1997, \aj, 113, 806

\bibitem[{{Gizis} {et~al.}(2002){Gizis}, {Reid}, \&
  {Hawley}}]{2002AJ....123.3356G}
{Gizis}, J.~E., {Reid}, I.~N., \& {Hawley}, S.~L. 2002, \aj, 123, 3356

\bibitem[{{Gregory}(2005)}]{2005ApJ...631.1198G}
{Gregory}, P.~C. 2005, \apj, 631, 1198

\bibitem[{{Griffin} {et~al.}(1976){Griffin}, {Radford}, {Harmer}, \&
  {Stickland}}]{1976Obs....96..153G}
{Griffin}, R.~F., {Radford}, G.~A., {Harmer}, D.~L., \& {Stickland}, D.~J.
  1976, The Observatory, 96, 153

\bibitem[{{Haario} {et~al.}(2001){Haario}, {Saksman}, \&
  {Tamminen}}]{haario2001}
{Haario}, P.~C., {Saksman}, E., \& {Tamminen}, J. 2001, Bernoulli, 7, 223

\bibitem[{{Hartman} {et~al.}(2010){Hartman}, {Bakos}, {Kov{\'a}cs}, \&
  {Noyes}}]{2010MNRAS.408..475H}
{Hartman}, J.~D., {Bakos}, G.~{\'A}., {Kov{\'a}cs}, G., \& {Noyes}, R.~W. 2010,
  \mnras, 408, 475

\bibitem[{{Hartman} {et~al.}(2011){Hartman}, {Bakos}, {Noyes}, {Sip{\H o}cz},
  {Kov{\'a}cs}, {Mazeh}, {Shporer}, \& {P{\'a}l}}]{2011AJ....141..166H}
{Hartman}, J.~D., {Bakos}, G.~{\'A}., {Noyes}, R.~W., {Sip{\H o}cz}, B.,
  {Kov{\'a}cs}, G., {Mazeh}, T., {Shporer}, A., \& {P{\'a}l}, A. 2011, \aj,
  141, 166

\bibitem[{{Hewett} {et~al.}(1985){Hewett}, {Irwin}, {Bunclark}, {Bridgeland},
  {Kibblewhite}, {He}, \& {Smith}}]{1985MNRAS.213..971H}
{Hewett}, P.~C., {Irwin}, M.~J., {Bunclark}, P., {Bridgeland}, M.~T.,
  {Kibblewhite}, E.~J., {He}, X.~T., \& {Smith}, M.~G. 1985, \mnras, 213, 971

\bibitem[{{Horne}(1986)}]{1986PASP...98..609H}
{Horne}, K. 1986, \pasp, 98, 609

\bibitem[{{Irwin} {et~al.}(2011){Irwin}, {Berta}, {Burke}, {Charbonneau},
  {Nutzman}, {West}, \& {Falco}}]{2011ApJ...727...56I}
{Irwin}, J., {Berta}, Z.~K., {Burke}, C.~J., {Charbonneau}, D., {Nutzman}, P.,
  {West}, A.~A., \& {Falco}, E.~E. 2011, \apj, 727, 56

\bibitem[{{Irwin} {et~al.}(2009{\natexlab{a}}){Irwin}, {Charbonneau},
  {Nutzman}, \& {Falco}}]{2009IAUS..253...37I}
{Irwin}, J., {Charbonneau}, D., {Nutzman}, P., \& {Falco}, E.
  2009{\natexlab{a}}, in IAU Symposium, Vol. 253, IAU Symposium, 37--43

\bibitem[{{Irwin} {et~al.}(2007){Irwin}, {Irwin}, {Aigrain}, {Hodgkin}, {Hebb},
  \& {Moraux}}]{2007MNRAS.375.1449I}
{Irwin}, J., {Irwin}, M., {Aigrain}, S., {Hodgkin}, S., {Hebb}, L., \&
  {Moraux}, E. 2007, \mnras, 375, 1449

\bibitem[{{Irwin} {et~al.}(2009{\natexlab{b}}){Irwin}, {Charbonneau}, {Berta},
  {Quinn}, {Latham}, {Torres}, {Blake}, {Burke}, {Esquerdo}, {F{\"u}r{\'e}sz},
  {Mink}, {Nutzman}, {Szentgyorgyi}, {Calkins}, {Falco}, {Bloom}, \&
  {Starr}}]{2009ApJ...701.1436I}
{Irwin}, J., {et~al.} 2009{\natexlab{b}}, \apj, 701, 1436

\bibitem[{{Johnson} \& {Soderblom}(1987)}]{1987AJ.....93..864J}
{Johnson}, D.~R.~H., \& {Soderblom}, D.~R. 1987, \aj, 93, 864

\bibitem[{{Kirkpatrick} {et~al.}(1991){Kirkpatrick}, {Henry}, \&
  {McCarthy}}]{1991ApJS...77..417K}
{Kirkpatrick}, J.~D., {Henry}, T.~J., \& {McCarthy}, Jr., D.~W. 1991, \apjs,
  77, 417

\bibitem[{{Kraus} \& {Hillenbrand}(2007)}]{2007AJ....134.2340K}
{Kraus}, A.~L., \& {Hillenbrand}, L.~A. 2007, \aj, 134, 2340

\bibitem[{{Kraus} {et~al.}(2011){Kraus}, {Tucker}, {Thompson}, {Craine}, \&
  {Hillenbrand}}]{2011ApJ...728...48K}
{Kraus}, A.~L., {Tucker}, R.~A., {Thompson}, M.~I., {Craine}, E.~R., \&
  {Hillenbrand}, L.~A. 2011, \apj, 728, 48

\bibitem[{{Kwee} \& {van Woerden}(1956)}]{1956BAN....12..327K}
{Kwee}, K.~K., \& {van Woerden}, H. 1956, \bain, 12, 327

\bibitem[{{Lacy}(1977)}]{1977ApJ...218..444L}
{Lacy}, C.~H. 1977, \apj, 218, 444

\bibitem[{{Landolt}(1992)}]{1992AJ....104..340L}
{Landolt}, A.~U. 1992, \aj, 104, 340

\bibitem[{{Leggett}(1992)}]{1992ApJS...82..351L}
{Leggett}, S.~K. 1992, \apjs, 82, 351

\bibitem[{{Leggett} {et~al.}(1996){Leggett}, {Allard}, {Berriman}, {Dahn}, \&
  {Hauschildt}}]{1996ApJS..104..117L}
{Leggett}, S.~K., {Allard}, F., {Berriman}, G., {Dahn}, C.~C., \& {Hauschildt},
  P.~H. 1996, \apjs, 104, 117

\bibitem[{{L{\'e}pine} \& {Shara}(2005)}]{2005AJ....129.1483L}
{L{\'e}pine}, S., \& {Shara}, M.~M. 2005, \aj, 129, 1483

\bibitem[{{Lopez-Morales}(2004)}]{2004PhDT........12L}
{Lopez-Morales}, M. 2004, PhD thesis, The University of North Carolina at
  Chapel Hill

\bibitem[{{L{\'o}pez-Morales}(2007)}]{2007ApJ...660..732L}
{L{\'o}pez-Morales}, M. 2007, \apj, 660, 732

\bibitem[{{L{\'o}pez-Morales} \& {Ribas}(2005)}]{2005ApJ...631.1120L}
{L{\'o}pez-Morales}, M., \& {Ribas}, I. 2005, \apj, 631, 1120

\bibitem[{{Lucy}(1967)}]{1967ZA.....65...89L}
{Lucy}, L.~B. 1967, \zap, 65, 89

\bibitem[{{Luhman}(1999)}]{1999ApJ...525..466L}
{Luhman}, K.~L. 1999, \apj, 525, 466

\bibitem[{{Luhman} \& {Rieke}(1998)}]{1998ApJ...497..354L}
{Luhman}, K.~L., \& {Rieke}, G.~H. 1998, \apj, 497, 354

\bibitem[{{MacDonald} \& {Mullan}(2011)}]{2011arXiv1106.1452M}
{MacDonald}, J., \& {Mullan}, D.~J. 2011, \apj, {submitted (arXiv:1106.1452)}

\bibitem[{{Metcalfe} {et~al.}(1996){Metcalfe}, {Mathieu}, {Latham}, \&
  {Torres}}]{1996ApJ...456..356M}
{Metcalfe}, T.~S., {Mathieu}, R.~D., {Latham}, D.~W., \& {Torres}, G. 1996,
  \apj, 456, 356

\bibitem[{{Morales} {et~al.}(2010){Morales}, {Gallardo}, {Ribas}, {Jordi},
  {Baraffe}, \& {Chabrier}}]{2010ApJ...718..502M}
{Morales}, J.~C., {Gallardo}, J., {Ribas}, I., {Jordi}, C., {Baraffe}, I., \&
  {Chabrier}, G. 2010, \apj, 718, 502

\bibitem[{{Morales} {et~al.}(2009){Morales}, {Ribas}, {Jordi}, {Torres},
  {Gallardo}, {Guinan}, {Charbonneau}, {Wolf}, {Latham}, {Anglada-Escud{\'e}},
  {Bradstreet}, {Everett}, {O'Donovan}, {Mandushev}, \&
  {Mathieu}}]{2009ApJ...691.1400M}
{Morales}, J.~C., {et~al.} 2009, \apj, 691, 1400

\bibitem[{{Mullan} \& {MacDonald}(2001)}]{2001ApJ...559..353M}
{Mullan}, D.~J., \& {MacDonald}, J. 2001, \apj, 559, 353

\bibitem[{{Nelson} \& {Davis}(1972)}]{1972ApJ...174..617N}
{Nelson}, B., \& {Davis}, W.~D. 1972, \apj, 174, 617

\bibitem[{{Nutzman} \& {Charbonneau}(2008)}]{2008PASP..120..317N}
{Nutzman}, P., \& {Charbonneau}, D. 2008, \pasp, 120, 317

\bibitem[{{O'Neal} {et~al.}(2004){O'Neal}, {Neff}, {Saar}, \&
  {Cuntz}}]{2004AJ....128.1802O}
{O'Neal}, D., {Neff}, J.~E., {Saar}, S.~H., \& {Cuntz}, M. 2004, \aj, 128, 1802

\bibitem[{{Popper} \& {Etzel}(1981)}]{1981AJ.....86..102P}
{Popper}, D.~M., \& {Etzel}, P.~B. 1981, \aj, 86, 102

\bibitem[{{Pr{\v s}a} {et~al.}(2011){Pr{\v s}a}, {Batalha}, {Slawson}, {Doyle},
  {Welsh}, {Orosz}, {Seager}, {Rucker}, {Mjaseth}, {Engle}, {Conroy},
  {Jenkins}, {Caldwell}, {Koch}, \& {Borucki}}]{2011AJ....141...83P}
{Pr{\v s}a}, A., {et~al.} 2011, \aj, 141, 83

\bibitem[{{Reid} {et~al.}(1995){Reid}, {Hawley}, \&
  {Gizis}}]{1995AJ....110.1838R}
{Reid}, I.~N., {Hawley}, S.~L., \& {Gizis}, J.~E. 1995, \aj, 110, 1838

\bibitem[{{Ribas}(2003)}]{2003A&A...398..239R}
{Ribas}, I. 2003, \aap, 398, 239

\bibitem[{{Ribas}(2006)}]{2006Ap&SS.304...89R}
---. 2006, \apss, 304, 89

\bibitem[{{Rockenfeller} {et~al.}(2006){Rockenfeller}, {Bailer-Jones}, \&
  {Mundt}}]{2006A&A...448.1111R}
{Rockenfeller}, B., {Bailer-Jones}, C.~A.~L., \& {Mundt}, R. 2006, \aap, 448,
  1111

\bibitem[{{Rodono} {et~al.}(1986){Rodono}, {Cutispoto}, {Pazzani}, {Catalano},
  {Byrne}, {Doyle}, {Butler}, {Andrews}, {Blanco}, {Marilli}, {Linsky},
  {Scaltriti}, {Busso}, {Cellino}, {Hopkins}, {Okazaki}, {Hayashi}, {Zeilik},
  {Helston}, {Henson}, {Smith}, \& {Simon}}]{1986A&A...165..135R}
{Rodono}, M., {et~al.} 1986, \aap, 165, 135

\bibitem[{{Shkolnik} {et~al.}(2010){Shkolnik}, {Hebb}, {Liu}, {Reid}, \&
  {Collier Cameron}}]{2010ApJ...716.1522S}
{Shkolnik}, E.~L., {Hebb}, L., {Liu}, M.~C., {Reid}, I.~N., \& {Collier
  Cameron}, A. 2010, \apj, 716, 1522

\bibitem[{{Southworth} {et~al.}(2004{\natexlab{a}}){Southworth}, {Maxted}, \&
  {Smalley}}]{2004MNRAS.351.1277S}
{Southworth}, J., {Maxted}, P.~F.~L., \& {Smalley}, B. 2004{\natexlab{a}},
  \mnras, 351, 1277

\bibitem[{{Southworth} {et~al.}(2004{\natexlab{b}}){Southworth}, {Zucker},
  {Maxted}, \& {Smalley}}]{2004MNRAS.355..986S}
{Southworth}, J., {Zucker}, S., {Maxted}, P.~F.~L., \& {Smalley}, B.
  2004{\natexlab{b}}, \mnras, 355, 986

\bibitem[{{Tegmark} {et~al.}(2004){Tegmark}, {Strauss}, {Blanton}, {Abazajian},
  {Dodelson}, {Sandvik}, {Wang}, {Weinberg}, {Zehavi}, {Bahcall}, {Hoyle},
  {Schlegel}, {Scoccimarro}, {Vogeley}, {Berlind}, {Budavari}, {Connolly},
  {Eisenstein}, {Finkbeiner}, {Frieman}, {Gunn}, {Hui}, {Jain}, {Johnston},
  {Kent}, {Lin}, {Nakajima}, {Nichol}, {Ostriker}, {Pope}, {Scranton},
  {Seljak}, {Sheth}, {Stebbins}, {Szalay}, {Szapudi}, {Xu}, {Annis},
  {Brinkmann}, {Burles}, {Castander}, {Csabai}, {Loveday}, {Doi}, {Fukugita},
  {Gillespie}, {Hennessy}, {Hogg}, {Ivezi{\'c}}, {Knapp}, {Lamb}, {Lee},
  {Lupton}, {McKay}, {Kunszt}, {Munn}, {O'Connell}, {Peoples}, {Pier},
  {Richmond}, {Rockosi}, {Schneider}, {Stoughton}, {Tucker}, {vanden Berk},
  {Yanny}, \& {York}}]{2004PhRvD..69j3501T}
{Tegmark}, M., {et~al.} 2004, \prd, 69, 103501

\bibitem[{{Terndrup} {et~al.}(2000){Terndrup}, {Stauffer}, {Pinsonneault},
  {Sills}, {Yuan}, {Jones}, {Fischer}, \&
  {Krishnamurthi}}]{2000AJ....119.1303T}
{Terndrup}, D.~M., {Stauffer}, J.~R., {Pinsonneault}, M.~H., {Sills}, A.,
  {Yuan}, Y., {Jones}, B.~F., {Fischer}, D., \& {Krishnamurthi}, A. 2000, \aj,
  119, 1303

\bibitem[{{Tody}(1993)}]{1993ASPC...52..173T}
{Tody}, D. 1993, in Astronomical Society of the Pacific Conference Series,
  Vol.~52, Astronomical Data Analysis Software and Systems II, ed.
  {R.~J.~Hanisch, R.~J.~V.~Brissenden, \& J.~Barnes} (San Francisco: ASP),
  173--+

\bibitem[{{Torres}(2007)}]{2007ApJ...671L..65T}
{Torres}, G. 2007, \apjl, 671, L65

\bibitem[{{Torres}(2010)}]{2010AJ....140.1158T}
---. 2010, \aj, 140, 1158

\bibitem[{{Torres} {et~al.}(2010){Torres}, {Andersen}, \&
  {Gim{\'e}nez}}]{2010A&ARv..18...67T}
{Torres}, G., {Andersen}, J., \& {Gim{\'e}nez}, A. 2010, \aapr, 18, 67

\bibitem[{{Torres} \& {Ribas}(2002)}]{2002ApJ...567.1140T}
{Torres}, G., \& {Ribas}, I. 2002, \apj, 567, 1140

\bibitem[{{Vaccaro} {et~al.}(2007){Vaccaro}, {Rudkin}, {Kawka}, {Vennes},
  {Oswalt}, {Silver}, {Wood}, \& {Smith}}]{2007ApJ...661.1112V}
{Vaccaro}, T.~R., {Rudkin}, M., {Kawka}, A., {Vennes}, S., {Oswalt}, T.~D.,
  {Silver}, I., {Wood}, M., \& {Smith}, J.~A. 2007, \apj, 661, 1112

\bibitem[{{van Hamme}(1993)}]{1993AJ....106.2096V}
{van Hamme}, W. 1993, \aj, 106, 2096

\bibitem[{{Viti} {et~al.}(2002){Viti}, {Jones}, {Maxted}, \&
  {Tennyson}}]{2002MNRAS.329..290V}
{Viti}, S., {Jones}, H.~R.~A., {Maxted}, P., \& {Tennyson}, J. 2002, \mnras,
  329, 290

\bibitem[{{Viti} {et~al.}(1997){Viti}, {Jones}, {Schweitzer}, {Allard},
  {Hauschildt}, {Tennyson}, {Miller}, \& {Longmore}}]{1997MNRAS.291..780V}
{Viti}, S., {Jones}, H.~R.~A., {Schweitzer}, A., {Allard}, F., {Hauschildt},
  P.~H., {Tennyson}, J., {Miller}, S., \& {Longmore}, A.~J. 1997, \mnras, 291,
  780

\bibitem[{{West} {et~al.}(2008){West}, {Hawley}, {Bochanski}, {Covey}, {Reid},
  {Dhital}, {Hilton}, \& {Masuda}}]{2008AJ....135..785W}
{West}, A.~A., {Hawley}, S.~L., {Bochanski}, J.~J., {Covey}, K.~R., {Reid},
  I.~N., {Dhital}, S., {Hilton}, E.~J., \& {Masuda}, M. 2008, \aj, 135, 785

\bibitem[{{West} {et~al.}(2005){West}, {Walkowicz}, \&
  {Hawley}}]{2005PASP..117..706W}
{West}, A.~A., {Walkowicz}, L.~M., \& {Hawley}, S.~L. 2005, \pasp, 117, 706

\bibitem[{{Wilson} \& {Devinney}(1971)}]{1971ApJ...166..605W}
{Wilson}, R.~E., \& {Devinney}, E.~J. 1971, \apj, 166, 605

\bibitem[{{Woolf} \& {Wallerstein}(2006)}]{2006PASP..118..218W}
{Woolf}, V.~M., \& {Wallerstein}, G. 2006, \pasp, 118, 218

\bibitem[{{Zahn}(1977)}]{1977A&A....57..383Z}
{Zahn}, J. 1977, \aap, 57, 383

\bibitem[{{Zucker} \& {Mazeh}(1994)}]{1994ApJ...420..806Z}
{Zucker}, S., \& {Mazeh}, T. 1994, \apj, 420, 806

\end{thebibliography}

\end{document}